\documentstyle[dina4,11pt]{article}
\begin{document}
\pagestyle{plain}
\newcommand{\be}{\begin{equation}}
\newcommand{\ee}{\end{equation}}
\newcommand{\bear}{\begin{eqnarray}}
\newcommand{\ear}{\end{eqnarray}}
\newcommand{\no}{\noindent}
\date{}
\renewcommand{\theequation}{\arabic{section}.\arabic{equation}}
\newcommand{\half}{{1\over 2}}
\newcommand{\e}{\mbox{e}}
\newcommand{\g}{\mbox{g}}
\renewcommand{\arraystretch}{2.5}
\newcommand{\GeV}{\mbox{GeV}}
\newcommand{\cL}{\cal L}
\newcommand{\D}{\cal D}
\newcommand{{\Tr}}{\rm Tr}
\newcommand{{\tr}}{\rm tr}
\newcommand{\Det}{\rm Det}
\newcommand{\PP}{\cal P}
\newcommand{\G}{{\cal G}}
\def\R{1\!\!{\rm R}}
\def\eins{  1\!{\rm l}  }
\newcommand{\symb}{\mbox{symb}}
\renewcommand{\arraystretch}{2.5}
\newcommand{\slD}{\raise.15ex\hbox{$/$}\kern-.57em\hbox{$D$}}

\pagestyle{empty}
\renewcommand{\thefootnote}{\fnsymbol{footnote}}
\topmargin =-.3in
\textheight=8.9in
\hskip 9cm {\sl IASSNS-HEP-96/90}
\vskip-.1pt 
\hskip 9cm {\sl DESY 96 - 225}
\vskip-.1pt
\hskip 9cm {\sl HD-THEP-96/17}
\vskip-.1pt
\hskip 9cm {\sl HUB-EP-96/13}
\vskip .4cm
\begin{center}
{\Large Constant External Fields in Gauge
Theory}\\
{\Large and the Spin 0,$\frac{1}{2}$,1
Path Integrals
} \\[1ex]
\vskip1.2cm
{\large Martin Reuter
}\\[1.5ex]
{\it Deutsches Elektronen-Synchrotron DESY\\
Notkestrasse 85, D-22603 Hamburg, Germany
\medskip\\}
\vskip.3cm
 {\large Michael G. Schmidt
}\\[1.5ex]
 {\it
  Institut f\"ur Theoretische Physik,
  Universit\"at Heidelberg\\
  Philosophenweg 16,
  D-69120 Heidelberg, Germany
\medskip\\}
\vskip.3cm
 {\large Christian Schubert
   \footnote{Supported by Deutsche Forschungsgemeinschaft}
}
\\[1.5ex]
{\it
School of Natural Sciences, 
Institute for Advanced Study\\
Olden Lane, Princeton, NJ 08540, USA
\medskip\\
}
and\\
 {\it
  Humboldt Universit\"at zu Berlin\\
 Invalidenstr. 110,
   D-10115 Berlin, Germany
\medskip \\
}
\vskip 1.5cm
 %
\vskip 1.5cm
 {\large \bf Abstract}
\end{center}
\begin{quotation}
\noindent
We investigate the usefulness of
the ``string-inspired technique''
for gauge theory calculations
in a constant external field background.
Our approach 
is based on Strassler's worldline
path integral approach to the
Bern-Kosower formalism, and on
the construction of
worldline (super--)
Green's functions incorporating 
external fields as well
as internal propagators.
The worldline path integral
representation 
of the gluon loop is
reexamined in detail.
We calculate
the two-loop effective
actions induced for a constant 
external field
by a scalar and spinor loop, 
and the corresponding
one-loop effective action
in the gluon loop case.
%
\end{quotation}
\clearpage
\renewcommand{\thefootnote}{\protect\arabic{footnote}}
\pagestyle{plain}

\setcounter{page}{1}
\setcounter{footnote}{0}

\topmargin=-.3in
\textheight=8.3in
\textwidth=7.2in

{\bf 1. Introduction}
\renewcommand{\theequation}{1.\arabic{equation}}
\setcounter{equation}{0}
\vskip10pt
It is by now well-established that techniques
from string perturbation theory can be used to
improve on calculational efficiency in ordinary
quantum field theory. The relevance of string theory 
for this purpose
is based on the fact
that many, and perhaps all, amplitudes in
quantum field theory can be represented as the
infinite string tension limits of appropriately
chosen (super--) string amplitudes. 
This is,
of course, an intrinsic and well-known property of 
string-theory. 

It is a more recent discovery, however, 
that such 
representations can lead to an interesting alternative
to standard Feynman diagram calculations.
Following earlier work on the $\beta$ -- function for
Yang-Mills-theory ~\cite{minahan,kaplunovsky,metseyt,bk88}, 
Bern and Kosower ~\cite{bk} considered
the infinite string tension limit 
of gauge boson scattering amplitudes,
formulated in an appropriate heterotic
string model.
By a detailed analysis of this limit, they
succeeded in deriving
a novel type of ``Feynman Rules''
for the construction of
ordinary one-loop gluon (photon) 
scattering amplitudes.
The resulting integral representations are
equivalent to the ones
originating from Feynman diagrams ~\cite{bd}, but 
lead
to a significant reduction of the number of terms to be 
computed in
gauge theory calculations. This property was then 
successfully
exploited to obtain both five gluon ~\cite{5glu} 
and four
graviton amplitudes ~\cite{4grav}.

Strassler ~\cite{str1} 
later showed that, 
for many cases of interest,
the same integral representations can be
derived from a first-quantized 
reformulation of
ordinary field theory.
In this approach, one 
starts with writing
the one-loop effective
action 
as a particle (super) path-integral,
of a type which has been known for
many years 
\cite{feyn,frad,cagolo,bdzdh,bdh,bssw,halsie,alvg,bardur}. 
Those path-integrals are then
considered as the field-theory analogues
of the Polyakov path integral, and
evaluated in a way analogous to string
perturbation theory (some suggestions
along similar lines had already been
made in \cite{polbook}).

\noindent
The resulting formalism offers an alternative to
standard field theory techniques which 
circumvents much of the apparatus of quantum field theory.
It works equally well for 
effective action and scattering
amplitude calculations, on-- and off--shell.
It has been applied to a number
of calculations in gauge theory
\cite{str2,ss1,fhss1,cdd,gussho}
and generalized to cases where,
besides gauge 
and scalar self-couplings,
Yukawa ~\cite{mnss1,dhogag1,holten} and
axial couplings ~\cite{mnss2,dhogag2,holten}
are present.

Due to its simplicity, it appears
also well-suited to the construction
of multiloop generalizations of
the Bern-Kosower formalism.
Steps in this direction have already
been taken by various authors, and along
different lines
~\cite{lam,roland,dlmmr}.
In particular, the original Bern-Kosower program
becomes very hard to carry through at the
two-loop level, due to the
complicated structure of
genus two string amplitudes.
Nevertheless, recently substantial progress 
has beeen achieved in this line of work
~\cite{dlmmr,rolsat}.

A multiloop generalization following the
spirit of Strassler's approach
has been proposed by two of the
present authors, first
for scalar field theories in ~\cite{ss2}. 
This generalization
uses the concept of worldline Green's
functions on graphs ~\cite{ss2,sato,rolsat}, 
and leads to
integral representations combining
whole classes of graphs.
This work was extended to QED in 
~\cite{ss3}, and its
practical viability demonstrated by 
a recalculation of the two-loop
(scalar and spinor) QED $\beta$
-- functions. 
For the case of multiloop
amplitudes in scalar QED,
a more comprehensive treatment
along the same lines
was given 
in ~\cite{dss}. This includes
amplitudes involving external scalars.

An important role in quantum electrodynamics
is played by
calculations involving constant
external fields. 
This subject originates with
Euler and Heisenberg's classic 
one--loop calculation of the
static limit of photon scattering 
in spinor
quantum electrodynamics ~\cite{eulhei}.
Schwinger's
introduction of the proper--time
method in 1951
~\cite{schwinger51} 
allowed him to reproduce this
result, and the analogous one for
scalar quantum electrodynamics,
with considerably less effort. 
For calculations of this type, it has turned out 
generally
advantageous to take account of the external
field already at the level of the Feynman rules,
i.e. to absorb it into the free propagators.
Appropriate formalisms have been developed both for
QED ~\cite{geheniau,tsai,bks} and QCD
~\cite{vzns}.

The purpose of the present paper is threefold.
First, we would like to cast the worldline
multiloop formalism proposed in ~\cite{ss2,ss3}
into a form suitable for constant external field
calculations in quantum electrodynamics.
This will be achieved in a way analogous to field
theory, namely by modifying the worldline Green's
functions so as to take the external field
into account.
At the one-loop level, similar proposals
have already been made by several authors ~\cite
{cdd,gussho,shaisultanov} (see also ~\cite{mckshe}).

Second, we will apply this formalism to
a two-loop effective action calculation
in QED.
In ~\cite{ss1} the worldline formalism
was used for a recalculation of
the Euler--Heisenberg Lagrangian.
In the present work, we will extend this calculation
to the two-loop level, 
and calculate the correction
to this effective Lagrangian due to
one internal photon, both for scalar and
spinor QED.
We discuss in how far this calculation improves
on previous calculations of the same quantities
~\cite{ritusscal,lebedev,ginzburg,ritusspin,ditreu}.
 
Third, we reconsider the 
super path integral representation
of the spin 1 particle, and use this
path integral for 
calculating the 
effective action induced for an external constant
pseudo--abelian field by a gluon loop in Yang-Mills-
theory.
While external gluons pose no 
particular problems in
the worldline formalism, apart from
forcing path--ordering, 
internal gluons are a much more
delicate matter. 
While it is not difficult
to construct free path integrals describing
particles of arbitrary spin, those
constructions usually run into
consistency problems if one tries to
couple a path integral with spin higher than
$1\over 2$ to
a spin-1 background ~\cite{bucshv,ggt}.
A somewhat non--standard path integral, 
mimicking the superstring, had been
proposed by Strassler 
~\cite{str1} in his rederivation of
the Bern-Kosower rules for the gluon loop case.
To the best of our knowledge, it has
not yet been proven that this path
integral correctly reproduces all the 
pinch terms implicit in the
Bern--Kosower master formula.

We will first give a rigorous
derivation
of a spin-1 path integral which,
while not identical with the one used
by Strassler, is easily seen to be 
equivalent. We then use it for
calculating the effective
action induced by a gluon loop for a constant
pseudo--abelian background gluon
field.
The result will again be in agreement with the
literature ~\cite{shore}, and provides a nontrivial check
on the correctness of Strassler's proposal.

The organization of this paper is as follows. In
chapter 2 we review the worldline-formalism
for one-loop photon scattering in scalar and spinor
QED, and its generalization to gluon scattering.
We then indicate the changes which are
necessary to take a constant external field into account.
Chapter 3 extends this analysis to the gluon loop.
This so far includes worldline calculations
of the one-loop effective actions induced for a
(pseudo--abelian) constant external field by
a scalar, spinor and gluon loop.
For the QED case, we then extend the constant field
formalism to the multiloop level in chapter 4.
We apply it to calculations of the corresponding
two--loop effective actions for scalar 
QED in chapter 5 and for spinor QED in chapter 6.
Chapter 7 contains some remarks on the various
ways of calculating the two--loop QED $\beta$ -- function
in this formalism.
Our results will be discussed in chapter 8.
In appendix A we discuss
the path integral representation of the electron
propagator in an external field.
In appendix B we derive the various 
worldline Green's functions
used in our calculations,
and discuss some of their properties. 
Appendix C contains a
collection of some results concerning the
determinants which we encounter in the evaluation
of the spin-1 path integral.

\vskip15pt

{\bf 2. One--Loop Photon Amplitudes in a Constant Background Field}
\renewcommand{\theequation}{2.\arabic{equation}}
\setcounter{equation}{0}
\vskip10pt

We begin with shortly reviewing how 
one--loop photon (gluon)
scattering amplitudes are calculated in
the worldline formalism ~\cite{str1,ss1,ss3}.
In his rederivation of the Bern--Kosower rules
for gauge boson scattering off a spinor loop,
Strassler sets out from the following well-known
path integral representation for the corresponding
one--loop effective action (see e.g.
~\cite{polbook,holten86}):

 \begin{eqnarray}
\Gamma\lbrack A\rbrack &  = &- 2 {\displaystyle\int_0^{\infty}}
{dT\over T}
e^{-m^2T}
{\displaystyle\int}_P 
{\cal D} x
{\displaystyle\int}_A 
{\cal D}\psi\nonumber\\
& \phantom{=}
&\times{\rm tr}{\cal P} 
{\rm exp}\biggl [- \int_0^T d\tau
\Bigl ({1\over 4}{\dot x}^2 + {1\over
2}\psi\dot\psi
+ ieA_{\mu}\dot x_{\mu} - ie
\psi_{\mu}F_{\mu\nu}\psi_{\nu}
\Bigr )\biggr ]\nonumber\\
\label{spinorpi}
\end{eqnarray}

\noindent
In this formula,
$T$ is the usual Schwinger proper--time parameter.
The periodic
functions $x^{\mu}(\tau )$ 
describe the embedding of the circle
with circumference $T$ into $D\;$-- dimensional
Euclidean spacetime, while
the $\psi^{\mu}(\tau )$'s are
antiperiodic Grassmann functions.
The periodicity properties are expressed
by the subscripts $P,A$ on the path integral.
The colour trace $tr$ and the path ordering
$\cal P$ apply, of course, only to the
nonabelian case. We have 
chosen a constant euclidean
worldline metric.

The case of a scalar loop is obtained simply
by discarding all Grassmann quantities and the
global factor of -2, which takes care of
the difference in
statistics and degrees of freedom.

Analogous path integral representations exist for the
scalar and electron propagators in a background field
~\cite{feyn,frad,bc,hht}. 
The integration is then over a space of
paths with fixed boundary conditions. 
Those 
are obvious in the scalar case,
while in the fermionic case there
has been some debate on the correct
boundary conditions
to be imposed
on the Grassmann path integral ~\cite{bc,hht,polbook}.
An adequate 
path integral representation for the gluon propagator
seems to be missing in the literature, and will
be derived in chapter 3. 
The more familiar cases of the scalar and 
spin-$\frac{1}{2}$
propagators are included 
in appendix A for completeness.

While in the present paper we will make
use only of the loop path integrals,
we expect 
those propagator 
path integral representations to
play an important role in future extensions of
this formalism. 

A number of
different techniques have been applied to
calculating this worldline 
path integral, and various generalizations
thereof 
(see e.g. 
~\cite
{hjs,rajeev,gitshv,bastianelli,bpsn,mck,dilmck,kks,gitzla,lunev}).
In the ``string--inspired'' approach,
the first step is to split the coordinate path integral into
center of mass and relative coordinates,

\begin{eqnarray}
{\displaystyle\int}{\cal D}x &=&
{\displaystyle\int}d x_0{\displaystyle\int}{\cal D} y\nonumber\\
x^{\mu}(\tau) &=& x^{\mu}_0  +
y^{\mu} (\tau )\nonumber\\
\int_0^T d\tau\,   y^{\mu} (\tau ) &=& 0\quad .\nonumber\\
\label{split}
\end{eqnarray}

\noindent
The path integrals over $y$ and $\psi$ 
are then evaluated by 
Wick contractions, as in a one-dimensional field theory 
on the circle. The Green's functions to be used are 

\begin{eqnarray}
\langle y^{\mu}(\tau_1)y^{\nu}(\tau_2)\rangle
   & = &- g^{\mu\nu}G_B(\tau_1,\tau_2)
    = \quad - g^{\mu\nu}\biggl[ \mid \tau_1 - \tau_2\mid -
{{(\tau_1 - \tau_2)}^2\over T}\biggr],\nonumber\\
\langle \psi^{\mu}(\tau_1)\psi^{\nu}(\tau_2)\rangle
   & = &{1\over 2}\, g^{\mu\nu} G_F(\tau_1,\tau_2)
   = \quad {1\over 2}\, g^{\mu\nu}{\rm sign}
(\tau_1 -\tau_2 )\, .\nonumber\\
\label{Green's}
\end{eqnarray}

\noindent
We will often
abbreviate $G_B(\tau_1,\tau_2)=: G_{B12}$ etc.They solve the differential equations

\begin{eqnarray}
{1\over 2}{\partial^2\over {\partial\tau_1}^2}
G_B(\tau_1,\tau_2) &=& \delta(\tau_1 -\tau_2) 
-{1\over T}\nonumber\\
{1\over 2}{\partial\over {\partial\tau_1}}
G_F(\tau_1,\tau_2) &=& 
\delta(\tau_1 -\tau_2) \nonumber\\
\label{Gdiffequ}
\end{eqnarray}
\noindent
with periodic (antiperiodic) boundary
conditions for $G_B$($G_F$).
Some freedom exists in the definition of $G_B$,
which has been discussed elsewhere ~\cite{fhss1,fss};
in particular, a constant added to $G_B$ would drop
out after momentum conservation is applied.

\noindent
With our conventions, the free path integrals are
normalized as

\begin{eqnarray}
{\displaystyle\int}_P {\cal D} y\,
 {\rm exp}\Bigl [- \int_0^T d\tau
{1\over 4}{\dot y}^2\Bigr ]
& = & {\lbrack 4\pi T\rbrack}^{-{D\over 2}}\nonumber\\
{\displaystyle\int}_A {\cal D} \psi\,
{\rm exp}\Bigl [- \int_0^T d\tau
{1\over2}\psi\dot\psi\Bigr ]
& = & 1\qquad .\nonumber\\
\label{norm}
\end{eqnarray}
\noindent
The result of this evaluation is the 
one-loop effective Lagrangian
${\cal L}(x_0)$. Combined with a 
covariant Taylor
expansion of the external field at 
$x_0$, this yields a new and 
highly efficient
algorithm for calculating 
higher derivative expansions in gauge
theory ~\cite{ss1,fhss1}.

One--loop scattering amplitudes are obtained
by specializing the background to 
a finite sum of plane waves of definite
polarization. Equivalently, one may define
integrated vertex operators 

\begin{equation}
V_A=
T^a
\int_0^T d\tau\,\bigl[\dot x_{\mu}\varepsilon_{\mu}
-2{\rm i}\psi_{\mu}\psi_{\nu}k_{\mu}\varepsilon_{\nu}\bigr]
{\rm exp}[ikx({\tau})]
\label{vertex}
\end{equation}

\noindent
for external photons (gluons) of definite momentum and
polarization, and calculate multiple insertions 
of those vertex operators
into the free path integral
(of course,
only the first term has to be taken for the
scalar loop).
$T^a$ denotes a generator of the gauge group
in the representation of the loop particle.
In the nonabelian case, for the spinor loop
an additional two-gluon vertex operator is
required ~\cite{str1}.

 The path
integrals are performed using the well-known
formulas for Wick--contractions involving
exponentials, e.g.

\begin{equation}
\Bigl\langle {\rm exp}{[ik_1\cdot x(\tau_1)]}
{\rm exp}{[ik_2\cdot x(\tau_2)]}\Bigr\rangle
= {\rm exp}{\Bigl[G_B(\tau_1,\tau_2)k_1\cdot k_2\Bigr]}
\label{expowick}
\end{equation}
\noindent
etc. (the factors $
\dot x_{\mu}\varepsilon_{\mu}$
may be formally exponentiated for convenience).

Explicit execution of the $\psi$ -- path integral
would be algebraically equivalent to the
calculation of the
corresponding Dirac traces in field theory.
It can be circumvented by the following remarkable
feature of the Bern-Kosower formalism, which may
be understood as a consequence
of worldsheet 
~\cite{berntasi}
or worldline ~\cite{str1} supersymmetry. 
After evaluation of the $x$ -- path integral,
one is left with an integral over the
parameters $T,\tau_1,\ldots,\tau_N$, where
$N$ is the number of external legs. 
In the nonabelian case, the path--ordering leads
to ordered $\tau$ -- integrals,
$\int\prod_{i=1}^{N-1}d\tau_i\theta(\tau_i-\tau_{i+1})$.
The integrand  
is an expression consisting of an exponential factor

$${\rm exp}\Bigl[\sum_{i<j}G_B(\tau_i,\tau_j) k_ik_j\Bigr]$$

\noindent
multiplied by a polynomial in the first and second
derivatives of $G_B$,

\begin{eqnarray}
\dot G_B(\tau_1,\tau_2) &=& {\rm sign}(\tau_1 - \tau_2)
- 2 {{(\tau_1 - \tau_2)}\over T}\nonumber\\
\ddot G_B(\tau_1,\tau_2)
&=& 2 {\delta}(\tau_1 - \tau_2)
- {2\over T}\quad \nonumber\\
\label{GdGdd}
\end{eqnarray}
\noindent
(here and in the following, a ``dot'' denotes
differentiation with respect to the first
variable). 

All $\ddot G_B$'s can 
be eliminated by 
partial integrations on the worldline,
leading to an equivalent parameter
integral dependent only on
$G_B$ and $\dot G_B$.
According to the Bern--Kosower rules,
all contributions
from fermionic Wick contractions 
may then 
be taken into account simply 
by simultaneously
replacing every closed 
cycle of $\dot G_B$'s appearing, say
$\dot G_{Bi_1i_2} 
\dot G_{Bi_2i_3} 
\cdots
\dot G_{Bi_ni_1}$, 
by its ``supersymmetrization'',

\begin{equation}
\dot G_{Bi_1i_2} 
\dot G_{Bi_2i_3} 
\cdots
\dot G_{Bi_ni_1}
\rightarrow 
\dot G_{Bi_1i_2} 
\dot G_{Bi_2i_3} 
\cdots
\dot G_{Bi_ni_1}
\nonumber\\
-
G_{Fi_1i_2}
G_{Fi_2i_3}
\cdots
G_{Fi_ni_1}.\nonumber\\
\label{subrule}
\end{equation}

\noindent
Note that an expression is considered
a cycle already if it
can be put into cycle form using the
antisymmetry of $\dot G_B$
(e.g. $\dot G_{Bab}\dot G_{Bab}
=-\dot G_{Bab}\dot G_{Bba}$). 
Unfortunately the practical value of this procedure
rapidly diminishes with increasing number of external
legs ($\ge 5$), 
as the number of terms making up the
integrand starts to significantly
increase in the partial integration
procedure ~\cite{bernpc}. 

The worldline 
supersymmetry (see eq. ({A.9})) 
makes it possible to
combine the $x$ -- and $\psi$ -- path
integrals into the following
super path 
integral ~\cite{bdzdh,polbook,rajeev,andtse,ss3}

\begin{equation}
\Gamma\lbrack A\rbrack   = - 2{\displaystyle\int_0^{\infty}}
{dT\over T}
e^{-m^2T}
{\displaystyle\int} {\cal D}\, X
{\rm tr}{\cal P}{\rm e}^{-
\int_0^T d\tau\int d\theta \, \Bigl [-{1\over 4}XD^3 X 
-{\rm ie}DX_{\mu}A_{\mu}(X)\Bigr ]},
\label{superaction}
\end{equation}

\noindent

\begin{eqnarray}
X^{\mu} &=& x^{\mu} 
+ \sqrt 2\,\theta\psi^{\mu} 
=x_0^{\mu}+Y^{\mu}\nonumber\\
D &=& {\partial\over{\partial\theta}} - 
   \theta
{\partial\over{\partial\tau}} \nonumber\\
\int d\theta\theta &=& 1\qquad\qquad . \nonumber\\
\label{defsuper}
\end{eqnarray}

\noindent
The photon (gluon) vertex operator is then rewritten as

\begin{equation}
-T^a\int_0^T d\tau d\theta \varepsilon_{\mu}
DX_{\mu}{\rm exp}[ikX],
\label{supervertex}
\end{equation}

\noindent
and we are left with a single Wick--contraction
rule

\begin{eqnarray}
\langle Y^{\mu}(\tau_1,\theta_1)
Y^{\nu}(\tau_2,\theta_2)\rangle
   & = &- g^{\mu\nu}\hat G(\tau_1,\theta_1;\tau_2,\theta_2)
\nonumber\\
\hat G(\tau_1,\theta_1;\tau_2,\theta_2)
&=& G_B(\tau_1,\tau_2) +
\theta_1\theta_2 G_F(\tau_1,\tau_2).\nonumber\\
\label{superpropagator}
\end{eqnarray}

\noindent
The fermion loop case can thus be made to look
formally identical to the scalar loop case, and be
regarded as its ``supersymmetrization''. This
analogy has its roots in the fact that the
string-inspired technique corresponds to the use
of a second-order formalism for fermions in field
theory (see ~\cite{morgan} and ref. therein),
instead of the usual first-order ones.
In practical terms it means that
the Grassmann Wick contractions are replaced
by a number of Grassmann integrals, which have
to be performed at a later stage of the
calculation.
Ultimately the superfield
formalism leads to the same collection of
parameter integrals to be performed, however
we have found it useful for keeping
intermediate expressions compact.
In the nonabelian case it has the further advantage
that the introduction of an additional
two--gluon vertex operator can be avoided.
Instead one introduces a suitable supersymmetric
generalization of the functions
$\theta(\tau_i - \tau_{i+1})$ appearing
in the ordered $\tau$ -- integrals ~\cite{andtse}.

Now let us assume that we have, in addition to the
background field $A^{\mu}(x)$ we started with,
a second one, $\bar A^{\mu}(x)$, 
with constant field strength tensor
$\bar F_{\mu\nu}$. 
We will restrict ourselves
to the abelian case for the
remainder of this chapter.
Using Fock--Schwinger gauge
centered at $x_0$, we may
take $\bar A^{\mu}(x)$ to be of the form

\begin{equation}
\bar A_{\mu}(x) = 
{1\over 2}y_{\nu}\bar F_{\nu\mu}\quad .\\
\label{fockschwinger}
\end{equation}

\noindent
The constant field contribution to the 
worldline lagrangian may then be written
as

\begin{equation}
\Delta{\cal L} = {1\over 2}iey_{\mu}\bar F_{\mu\nu}
\dot y_{\nu} - ie\psi_{\mu}\bar F_{\mu\nu}\psi_{\nu}\\
\label{DelatLkomp}
\end{equation}
\noindent
in components, or as

\begin{equation}
\Delta{\cal L} = -{1\over 2}ieY_{\mu}\bar F_{\mu\nu}
DY_{\nu}\\
\label{DeltaLsuper}
\end{equation}
\noindent
in the superfield formalism.

In any case, it is still quadratic in the worldline
fields, and therefore need not be considered part
of the interaction lagrangian; we can absorb it
into the free worldline propagator(s). 
This means that we need to solve, instead of
eqs. (~\ref{Gdiffequ}) for the worldline
Green's functions, the modified equations

\begin{eqnarray}
{1\over 2}\biggl(
{\partial^2\over {\partial\tau_1}^2}
-2ie{\bar F}{\partial\over {\partial\tau_1}}
\biggr){\cal G}_{B}(\tau_1,\tau_2) 
&=& \delta(\tau_1 -\tau_2) 
-{1\over T}\\
{1\over 2}\biggl(
{\partial\over {\partial\tau_1}}
-2ie{\bar F}\biggr)
{\cal G}_{F}(\tau_1,\tau_2) &=& 
\delta(\tau_1 -\tau_2) \\
\label{GdiffequwithF}\nonumber
\end{eqnarray}

\noindent
These equations will be solved in appendix B,
with the result (deleting the ``bar'' again)

\begin{eqnarray}
{\cal G}_{B}(\tau_1,\tau_2) &=&
{1\over 2{(eF)}^2}\biggl({eF\over{{\rm sin}(eFT)}}
{\rm e}^{-ieFT\dot G_{B12}}
+ieF\dot G_{B12} -{1\over T}\biggr)
\nonumber\\
{\cal G}_{F}(\tau_1,\tau_2) &=&
G_{F12}
{{\rm e}^{-ieFT\dot G_{B12}}\over {\rm cos}(eFT)}
\nonumber\\
\label{calGBGF}
\end{eqnarray}
\noindent
Equivalent expressions have
been given for the pure magnetic field case
in ~\cite{cdd}, and for the general case in
~\cite{shaisultanov}.
These expressions should be understood as power
series in the field strength matrix $F$
(note that eqs.(~\ref{calGBGF})
do {\sl not} assume 
invertibility of $F$).
Note also that the generalized Green's functions are still
translation invariant in $\tau$, and thus
functions of $\tau_1 - \tau_2$.
They are, in general, nontrivial Lorentz matrices,
so that the Wick contraction rules eq.(\ref{Green's})
have to be rewritten as

\begin{eqnarray}
\langle y^{\mu}(\tau_1)y^{\nu}(\tau_2)\rangle
&=&
-{\cal G}_B^{\mu\nu}(\tau_1,\tau_2),\nonumber\\
\langle\psi^{\mu}(\tau_1)\psi^{\nu}(\tau_2)\rangle
&=&
\frac{1}{2}{\cal G}_F^{\mu\nu}(\tau_1,\tau_2).\nonumber\\
\label{exfieldGreen's}
\end{eqnarray}
\noindent
Again momentum conservation leads to the freedom
to subtract from
${\cal G}_B$ its
constant coincidence limit,

\begin{equation}
{\cal G}_{B}(\tau,\tau)=
{1\over 2{(eF)}^2}
\biggl(eF\cot(eFT)-{1\over T}
\biggr)
\label{coincidence1}
\end{equation}
\noindent
To correctly obtain this and other coincidence
limits, one has to apply the rules

\begin{equation}
\dot G_B(\tau,\tau)=0,\quad
\dot G_B^2(\tau,\tau)=1.
\label{coincidencerules}
\end{equation}
\noindent
More generally, coincidence limits should always be taken
{\sl after} derivatives.

\noindent
We also need the
generalizations of $\dot G_B,\ddot G_B$,
which turn out to be

\begin{eqnarray}
\dot{\cal G}_B(\tau_1,\tau_2)
&\equiv&2\langle\tau_1\mid
{\bigl(\partial_{\tau} -2ieF\bigr)}^{-1}
\mid\tau_2\rangle
=
{i\over eF}\biggl({eF\over{{\rm sin}(eFT)}}
{\rm e}^{-ieFT\dot G_{B12}}-{1\over T}\biggr)
\nonumber\\
\ddot{\cal G}_{B}(\tau_1,\tau_2)
&\equiv&2\langle\tau_1\mid
{\bigl({\bf I}-2ieF{\partial_{\tau}}^{-1}\bigr)}
^{-1}\mid\tau_2\rangle
= 2\delta_{12} -2{eF\over{{\rm sin}(eFT)}}
{\rm e}^{-ieFT\dot G_{B12}}\nonumber\\
\label{derivcalGB}
\end{eqnarray}
\noindent
Let us also give the first few terms of the Taylor
expansions in $F$ for those four functions:

\begin{eqnarray}
{\cal G}_{B12} &=& G_{B12}-{T\over 6}
-{i\over 3}
\dot G_{B12}G_{B12}TeF+({T\over 3}G_{B12}^2
-{T^3\over 90}){(eF)}^2+O(F^3)\nonumber\\
\dot{\cal G}_{B12} 
&=&\dot G_{B12}+2i\bigl(G_{B12}-{T\over 6}\bigr)eF
+{2\over 3}\dot G_{B12}G_{B12}T{(eF)}^2 +  O(F^3)
\nonumber\\
\ddot{\cal G}_{B12} 
&=& \ddot G_{B12}+2i\dot G_{B12}eF
-4\bigl(G_{B12}-{T\over 6}\bigr){(eF)}^2+O(F^3)\nonumber\\
{\cal G}_{F12}&=& G_{F12}-iG_{F12}\dot G_{B12}TeF
+2G_{F12}G_{B12}T{(eF)}^2+O(F^3)\nonumber\\
\label{GB(F)expand}
\end{eqnarray}
\noindent
Again ${\cal G}_B$ and ${\cal G}_F$ may be assembled into
a superpropagator,

\begin{equation}
\hat {\cal G}(\tau_1,\theta_1;\tau_2,\theta_2)
\equiv {\cal G}_B(\tau_1,\tau_2) +
\theta_1\theta_2 {\cal G}_F(\tau_1,\tau_2).\\
\label{calsuperpropagator}\nonumber
\end{equation}
\noindent
At first sight this definition would
seem not to accomodate the
nonvanishing
coincidence limit of ${\cal G}_F$
(which can {\sl not} be subtracted).
Nevertheless, 
comparison with the component field formalism shows
that the correct expressions are again
reproduced if one takes coincidence limits
{\sl after} superderivatives. For instance,
the correlator 
$\langle D_1X(\tau_1,\theta_1)X
(\tau_1,\theta_1)\rangle$
is evaluated by calculating  

\begin{equation}
\langle D_1X(\tau_1,\theta_1)
X(\tau_2,\theta_2)\rangle
=
\theta_1\dot{\cal G}_{B12}-\theta_2{\cal G}_{F12},
\label{wickDXX}
\end{equation}
\noindent
and then setting $\tau_2=\tau_1$.

It is easily seen that the substitution
rule eq.(~\ref{subrule}) continues to hold, if one
defines the cycle property solely in terms of the
$\tau$ -- indices, irrespectively of what happens
to the Lorentz indices. For example, the expression

\begin{equation}
\varepsilon_1\dot{\cal G}_{B12}k_2\,
\varepsilon_2\dot{\cal G}_{B23}\varepsilon_3\,
k_3\dot{\cal G}_{B31}k_1\,
\label{expression}
\end{equation}
\noindent
would have to be replaced by

\begin{equation}
\varepsilon_1\dot{\cal G}_{B12}k_2\,
\varepsilon_2\dot{\cal G}_{B23}\varepsilon_3\,
k_3\dot{\cal G}_{B31}k_1\,
-
\varepsilon_1{\cal G}_{F12}k_2\,
\varepsilon_2{\cal G}_{F23}\varepsilon_3\,
k_3{\cal G}_{F31}k_1\quad .
\label{Fsubrule}
\end{equation}
\noindent
The only novelty is again due to the fact that,
in contrast to $\dot G_B$ and
$G_F$, $\dot {\cal G}_B$ and ${\cal G}_F$ have
non-vanishing coincidence limits,

\begin{eqnarray}
\dot {\cal G}_B(\tau,\tau) &=& i{\rm cot}(eFT)
-{i\over eFT}\\
{\cal G}_F(\tau,\tau) &=& -i\,{\rm tan}(eFT)\\
\label{coincidence2}\nonumber
\end{eqnarray}
\noindent
As a consequence, we now have also a substitution rule for
one--cycles,

\begin{equation}
\dot{\cal G}_B(\tau_i,\tau_i)\rightarrow
\dot{\cal G}_B(\tau_i,\tau_i)
-{\cal G}_F(\tau_i,\tau_i)
= {i\over {\rm sin}(eFT){\rm cos}(eFT)}
- {i\over eFT}
\quad .
\label{onecycle}
\end{equation}

This is almost all we need to know for computing one-loop
photon scattering amplitudes, or the corresponding
effective action, in a constant overall background field. 
The only further information required at the one--loop
level is the change in the path integral determinants
due to the external field, i.e. the
vacuum amplitude in the constant field.
For spinor QED, this
just corresponds to the Euler--Heisenberg Lagrangian,
and has, in the present formalism,
been calculated in ~\cite{ss1}.
Let us
shortly retrace this calculation
(the fact that the Euler-Heisenberg integrand
may be represented as a superdeterminant
was already noted in ~\cite{ckmw}).
In the scalar QED case, we have to replace

\begin{equation}
{\displaystyle\int} {\cal D} y\,
 {\rm exp}\Bigl [- \int_0^T d\tau
{1\over 4}{\dot y}^2\Bigr ]
={\rm Det'}^{-{1\over 2}}_P\bigl[ -{\partial}^2_{\tau}\bigr]
=  {\lbrack 4\pi T\rbrack}^{-{D\over 2}}\\
\label{bosnorm}
\end{equation}
\noindent
by
\begin{equation}
{\rm Det'}^{-{1\over 2}}_P
\Bigl[ -{\partial}^2_{\tau}
+2ieF{\partial}_{\tau}\Bigr]
={\lbrack 4\pi T\rbrack}^{-{D\over 2}}
{\rm Det'}_P^{-{1\over 2}}
\Bigl[{\bf I} -2ieF{{\partial}_{\tau}}^{-1}\Bigr]
\quad 
\label{Fbosnorm}
\end{equation}
\noindent
(as usual, the prime denotes the absence of the
zero mode in a determinant).
Application of the 
$ln\,det = tr\,ln$ identity yields

\begin{eqnarray}
{\rm Det'}_P^{-{1\over 2}}
\Bigl[{\bf I} -2ieF{{\partial}_{\tau}}^{-1}\Bigr]
&=&
{\rm exp}\biggl[
-\frac{1}{2}\sum_{n=1}^{\infty}
{{(-1)}^{n+1}\over n}{(-2ie)}^n
{\rm tr}[F^n]{\rm Tr}[{\partial}_{\tau}^{-n}]
\biggr]
\nonumber\\
&=&
{\rm exp}\biggl[-
\frac{1}{2}\sum_{n=2\atop n\,{\rm even}}^{\infty}
{B_n\over n!n}{(2ieT)}^n
{\rm tr}[F^n]
\biggr]
\nonumber\\
&=&
{\rm det}^{-{1\over 2}}
\biggl[{{\rm sin}(eFT)\over eFT}\biggr],
\\
\label{scaldetcomp}\nonumber
\end{eqnarray}
\noindent
where the $B_n$ are the Bernoulli numbers.
In the second step eq.(\ref{B.23})
was used.
The analogous calculation
for the Grassmann path integral 
yields a factor

\begin{equation}
{\rm Det}_A^{+{1\over 2}}
\Bigl[{\bf I} -2ieF{{\partial}}^{-1}\Bigr]
=
{\rm det}^{1\over2}
\Bigl[\cos(eFT)\Bigr]
\; .
\label{grassfermcomp}
\end{equation}
For spinor QED we therefore find a 
total overall determinant factor of

\begin{equation}
{\lbrack 4\pi T\rbrack}^{-{D\over 2}}
{\rm det}^{-{1\over 2}}
\biggl[{{\rm tan}(eFT)\over eFT}
\biggr]
\quad .\\
\label{Ffermnorm}
\end{equation}
\noindent
Expressing ${\rm tr} [F^{2n}]$ in terms of the two invariants
of the electromagnetic field,

\begin{eqnarray}
{\rm tr} [F^{2n}]&=&
2\Bigl[
{(a^2)}^n+{(-b^2)}^n
\Bigr],\nonumber\\
a^2&=&\frac{1}{2}
\biggl[
{\bf E}^2-{\bf B}^2
+\sqrt
{{({\bf E}^2-{\bf B}^2)}^2
+4{({\bf E}\cdot
{\bf B})}^2
}
\biggr],\nonumber\\
b^2&=&\frac{1}{2}
\biggl[
-({\bf E}^2-{\bf B}^2)
+\sqrt
{{({\bf E}^2-{\bf B}^2)}^2
+4{({\bf E}\cdot
{\bf B})}^2
}
\biggr],\nonumber\\
\label{trafoFEB}
\end{eqnarray}
\noindent
we obtain the standard 
Schwinger proper-time representation
of the (unsubtracted) Euler-Heisenberg-Schwinger
 Lagrangians,

\begin{eqnarray}
{\cal L}_{\rm scal}^{(1)}[F]&=&
{1\over {(4\pi)}^2}\int_0^{\infty}
{dT\over T^3}
{\rm e}^{-m^2T}
{(eaT)(ebT)\over\sin(eaT)\sinh(ebT)},
\label{eulheiscal}\\
{\cal L}_{\rm spin}^{(1)}[F]&=&
-{2\over {(4\pi)}^2}\int_0^{\infty}
{dT\over T^3}
{\rm e}^{-m^2T}
{(eaT)(ebT)
\over\tan(eaT)\tanh(ebT)}\\
\label{eulheispin}\nonumber
\end{eqnarray}
\no


{\bf 3. Gauge Boson Loops in External Fields}
\renewcommand{\theequation}{3.\arabic{equation}}
\setcounter{equation}{0}
\vskip10pt

In this section we first express the one-loop effective action
and also the propagator of spin-1
gauge bosons  in an arbitrary external Yang-Mills field in terms of a
worldline path integral.
We then will evaluate the action in a covariantly constant background.

We employ the background gauge fixing technique
so that the effective action $\Gamma[A^a_\mu]$ becomes a gauge invariant
functional of $A^a_\mu$ \cite{abb,dr2}.
The gauge fixed classical action reads, in $D$ dimensions,
\be\label{3.1}
S[a;A]=\frac{1}{4} \int d^D x F^a_{\mu\nu} (A+a) F^a_{\mu\nu} (A+a)
+\frac{1}{2\alpha}\int d^D x\left( D^{ab}_\mu [A]a^b_\mu\right)^2.\ee
A priori, the background field $ A^a_\mu$ is unrelated to the
quantum field $a^a_\mu$. The kinetic operator of the gauge boson
fluctuations  one obtains as the second functional
derivative of $S[a,A]$
with respect to $a^a_\mu$,  at fixed $A^a_\mu$.  This leads to  the inverse  
propagator
\be\label{3.2}
{\cal D}^{ab}_{\mu\nu}=- D^{ac}_\rho D^{cb}_\rho \delta_{\mu\nu}-2ig
F^{ab}_{\mu\nu}\ee
and the effective action
\bear\label{3.3}
\Gamma[A]&=&\frac{1}{2}\ln \det({\cal D})\nonumber\\
&=&-\frac{1}{2}\int^{\infty}_0\frac{dT}{T} {\Tr}[e^{-T{\D}}].\ear
In writing down eq. (\ref{3.2}) we adopted the Feynman gauge $\alpha=1$.
The covariant derivative $D_\mu\equiv \partial_\mu+ig A^a_\mu  T^a$ and
the field strength $F^{ab}_{\mu\nu}\equiv F^c_{\mu\nu}(T^c)^{ab}$ are
matrices in the adjoint representation of the gauge group, with the
generators given by
$(T^a)^{bc}=-i f^{abc}$. The full effective action is obtained by
adding the contribution of the Faddeev-Popov ghosts to eq. (\ref{3.3}).
The evaluation of the ghost determinant proceeds along the same lines as
scalar QED and we shall not discuss it here.

In order to derive a path integral representation of the  heat-kernel

\begin{equation}
{\Tr}[\exp(-T{\D})]  
\end{equation}
\noindent
we first look at a slightly more general problem. We
generalize the operator ${\D}$ to
\be\label{3.4}
\widehat h_{\mu\nu}\equiv - D^2 \delta_{\mu\nu}+M_{\mu\nu}(x)\ee
where $M_{\mu\nu}(x)$ is an arbitrary matrix in color space. In
particular,
we do not assume that the Lorentz trace $M_{\mu\mu}$ is zero. Given
$M_{\mu\nu}$, we construct the following one-particle Hamilton operator:
\be\label{3.5}
\widehat H=(\widehat p_\mu+g A_\mu(\widehat x))^2-:
\widehat{\bar\psi}_\mu
M_{\nu\mu}\widehat\psi_\nu:\ee
The system under consideration has a graded phase-space coordinatized
by $x_\mu,p_\mu$ and two sets of anticommuting variables, $\psi_\mu$
and $\bar\psi_\mu$, which obey canonical anticommutation relations:
\be\label{3.6}
\widehat\psi_\mu \widehat{\bar\psi}_\nu+
\widehat{\bar\psi}_\nu\widehat\psi_\mu
=\delta_{\mu\nu}.\ee
For a reason which will become obvious in a moment we have adopted
the ``anti-Wick'' ordering in (\ref{3.5}): all $\bar\psi$'s are
on the right of all $\psi$'s, e.g.
\bear\label{3.7}
:\widehat\psi_\alpha\widehat{\bar\psi}_\beta:&=&\widehat\psi_\alpha
\widehat{\bar\psi}_\beta\nonumber\\
:\widehat{\bar\psi}_\beta \widehat\psi_\alpha:&=&-\widehat\psi_\alpha
\widehat{\bar\psi}_\beta.\ear
We can represent the commutation relations on a space of wave functions
$\Phi(x,\psi)$ depending on $x_\mu$ and a set of classical Grassmann
variables $\psi_\mu$. The ``position'' operators $\widehat x_\mu=x_\mu,\
\widehat\psi_\mu=\psi_\mu$ act multiplicatively on $\Phi$ 
and the conjugate
momenta as derivatives $\widehat p_\mu=-i\partial_\mu$ 
and $\widehat{\bar\psi}_\mu
=\partial/\partial\psi_\mu$. Thus the Hamiltonian becomes~\cite{met}
\be\label{3.8}
\widehat H=-D^2+\psi_\nu M_{\nu\mu}(x)\frac{\partial}
{\partial \psi_\mu}.\ee
The wave functions $\Phi$ have a decomposition of the form
\be\label{3.9}
\Phi(x,\psi)=\sum^D_{p=0}\frac{1}{p!}\ \phi^{(p)}_{\mu_1\dots \mu_p}
(x)\ \psi_{\mu_1}\dots\psi_{\mu_p}.\ee
This suggests the interpretation of $\Phi$ as an inhomogeneous
differential form on  $\R^D$ with the fermions $\psi_\mu$ playing
the role of the differentials $dx^\mu$~\cite{cm}. The form-degree
or, equivalently, the fermion number is measured  by the operator
\be\label{3.10}
\widehat F=\widehat\psi_\mu\widehat{\bar\psi}_\mu
=\psi_\mu\frac{\partial}{\partial
\psi_\mu}.\ee
We are particularly interested in one-forms:
\be\label{3.11}
\Phi(x,\psi)=\varphi_\mu(x)\psi_\mu.\ee
The Hamiltonian (\ref{3.8}) acts on them according to
\be\label{3.12}
(\widehat H\Phi)(x,\psi)=(\widehat h_{\mu\nu}\varphi_\nu)\psi_\mu.\ee
We see that, when restricted to the one-form sector,
 the quantum system with the Hamiltonian (\ref{3.5})
is equivalent to the one defined by the bosonic matrix Hamiltonian
$\widehat h_{\mu\nu}$~\cite{cm,met}.

The euclidean proper time
evolution of the wave functions $\Phi$ is implemented by the kernel
\be\label{3.13}
K(x_2,\psi_2, t_2|x_1,\psi_1,t_1)=\langle x_2,\psi_2| e^{-(t_2-t_1)
\widehat H}
|x_1,\psi_1\rangle\ee
which obeys the Schr\"odinger equation
\be\label{3.14}
\left(\frac{\partial}{\partial T}+\widehat H\right)
K(x,\psi,T|x_0,\psi_0,0)=0\ee
with the initial condition $K(x,\psi,0|x_0,\psi_0,0)=\delta(x-x_0)\delta
(\psi-\psi_0)$. It is easy to write down a path integral solution
to eq. (\ref{3.14}). For the trace of $K$ one obtains
\be\label{3.15}
{{\Tr}}[e^{-T\widehat H_W}]=\int_{\PP}{{\D}}  
x(\tau)\int_A{{\D}}\psi(\tau){{\D}}\bar\psi(\tau)
{\rm tr} {\PP} e^{-\int^T_0 d\tau L}\ee
with
\be\label{3.16}
L=\frac{1}{4}\dot x^2+ig A_\mu(x)\dot x_\mu+\bar\psi_\mu
[\partial_\tau \delta_{\mu\nu}-M_{\nu\mu}]\psi_\nu.\ee
We have again periodic boundary conditions for $x^{\mu}(\tau)$,
and antiperiodic boundary conditions for $\psi^{\mu}(\tau)$.
If we use periodic boundary conditions for the fermions we arrive at a
representation of the Witten index 
~\cite{witten} rather than the partition function:
\be\label{3.17}
{{\Tr}}[(-1)^{\widehat F} e^{-T\widehat H_W}]=\int_{\PP}{{\D}} x(\tau)
\int_{\PP}{{\D}}\psi(\tau){{\D}}\bar\psi(\tau){\rm tr} {\PP}
e^{-\int^T_0 d\tau L}.\ee
At this point we have to mention a subtlety which is frequently overlooked
but will be important later on. If we regard the Hamiltonian (\ref{3.5})
as
a function of the anticommuting $c$-numbers $\psi_\mu$ and $\bar\psi_\mu$
it is related to the classical Lagrangian (\ref{3.16}) by a standard
Legendre transformation. The information about the operator ordering
is implicit in the discretization which is used for the definition of the
path-integral. Different operator orderings correspond to different
discretizations. Here we shall adopt the midpoint prescription~\cite{sak}
 for the discretization, because only in this case the familiar
path-integral manipulations are allowed~\cite{sato2}.
 It is known~\cite{sak,sato2,ber,mizrahi,gj} that, at the 
operatorial level, this is
equivalent to using the Weyl ordered Hamiltonian $\widehat H_W$.  This is
the reason why we wrote $\widehat H_W$ rather than $\widehat H$ on the
l.h.s. of eqs.
(\ref{3.15}) and (\ref{3.17}). In order
  to arrive at the relation (\ref{3.12})
we had to assume that the fermion operators in $\widehat H$ 
are ``anti-Wick''
ordered. Weyl ordering amounts to a symmetrization in $\bar\psi$ and
$\psi$
so that
\bear\label{3.18}
\widehat H_W&=&(\widehat p_\mu+g A_\mu(\widehat x))^2+\frac{1}{2}
M_{\nu\mu}(\widehat x)(\widehat\psi_\nu\widehat{\bar\psi}_\mu-
\widehat{\bar\psi}_\mu\widehat\psi_\nu)\nonumber\\
&=&\widehat H-\frac{1}{2} M_{\mu\mu} (\widehat x).\ear
In the second line of (\ref{3.18}) we used (\ref{3.5})
and (\ref{3.6}). (With respect  to $\widehat x_\mu$ and $\widehat p_\mu$,
Weyl ordering is used throughout.) If we employ (\ref{3.18}) in
(\ref{3.15})
we obtain the following representation for the partition function of
the anti-Wick ordered exponential:
\bear\label{3.19}
{{\Tr}}[e^{-T\widehat H}]&=&\int_P {\D} x (\tau)\int_A{\D}\psi(\tau)
{\D}\bar\psi(\tau)\nonumber\\
&&\times {\rm tr} {\cal P}\exp\left[-\int^T_0 d\tau\left\{ L(\tau)
+\frac{1}{2} M_{\mu\mu}(x(\tau))\right\}\right].\ear
Let us now calculate the partition function ${{\Tr}}[\exp (-T\widehat h)]$
which is a generalization of the heat-kernel needed in eq. (\ref{3.3}).
By virtue of eq. (\ref{3.12}) we may write
\be\label{3.20}
{{\Tr}}[e^{-T\widehat h}]={{\Tr}}_1[e^{-T\widehat H}]\ee
where ``${\Tr}_1$'' denotes the trace in the one-form sector of the
theory which contains the worldline fermions. In order to perform
the projection on the one-form sector we identify $M_{\mu\nu}$ with
\be\label{3.21}
M_{\mu\nu}=C\delta_{\mu\nu}-2ig F_{\mu\nu}\ee
where $C$ is a real constant. As a consequence,
\be\label{3.22}
\widehat H=\widehat H_0+C\widehat F\ee
with
\be\label{3.23}
\widehat H_0\equiv(\widehat p_\mu+g A_\mu(\widehat x))^2-2ig F_{\nu\mu}
(\widehat x) \widehat\psi_\nu \widehat{\bar\psi}_\mu\ee
denoting the Hamiltonian which corresponds to the inverse propagator ${\D}$.
The fermion number operator $\widehat F\equiv\widehat\psi_\mu  
\widehat{\bar\psi}_\mu$
is anti-Wick ordered by definition. Its spectrum  consists of the
integers
$p=0,1,2,\dots D$. Note that $M_{\mu\mu}=DC$, and that because of the
antisymmetry of $F_{\mu\nu}$ the Hamiltonian $\widehat H_0$ has no ordering
ambiguity  in its fermionic piece.
It will prove useful to apply eq. (\ref{3.19}) not to $\widehat H$ directly,
but to
$\widehat H-C=\widehat H_0+C(\widehat F-1)$.
The result reads then
\bear\label{3.24}
&&{{\Tr}} [e^{-CT(\widehat F-1)}e^{-T\widehat H_0}]\nonumber\\
&&=\exp[-CT(\frac{D}{2}-1)]\int_P{\cal D}x(\tau)\int_A
{\cal D}\psi(\tau){\cal D}\bar\psi(\tau){\rm tr} {\PP}e^{-\int^T_0d\tau L}
\ear
with
\be\label{3.25}
L=\frac{1}{4}\dot x^2+ig A_\mu(x)\dot x_\mu
+\bar\psi_\mu[(\partial_\tau-C)\delta_{\mu\nu}-2igF_{\mu\nu}]
\psi_\nu\ee
After having performed the path integration in (\ref{3.24})
we shall send $C$ to infinity. While this has no effect in
the one-form sector, it leads to an exponential suppression
factor $\exp[-CT(p-1)]$ in the sectors with fermion
numbers $p=2,3,...D$. Hence only the zero and the one forms
survive the limit $C\to\infty$. In order to eliminate the
contribution from the zero forms we insert the projector
$[1-(-1)^{\widehat F}]/2$ into the trace. It projects on the
subspace of odd form degrees, and it is easily implemented
by combining periodic and antiperiodic boundary conditions
for $\psi_\mu$. In this manner we arrive at a representation
of the partition function of $\widehat H_0$ restricted to the
one-form sector:
\bear\label{3.26}
{{\Tr}}_1[e^{-T\widehat H_0}]
&=&\lim_{C\to\infty}{{\Tr}}\left[\frac{1}{2}(1-(-1)^{\widehat F})
e^{-CT(\widehat F-1)}e^{_-T\widehat H_0}\right]\nonumber\\
&=&\lim_{C\to\infty}\exp\left[-CT\left(\frac{D}{2}-1\right)
\right]\int_P{\cal D}x(\tau)\nonumber\\
&&\ \times\frac{1}{2}(\int_A-\int_P){{\D}}\psi(\tau){{\D}}\bar\psi(\tau)
{{\Tr}}{\PP}e^{-\int_0^Td\tau L}\ear
Because ${{\Tr}}[\exp(-T{{\D}})]={{\Tr}}_1[\exp(-T\widehat H_0)]$,
eq. (\ref{3.26}) implies for the effective action
\bear\label{3.27}
\Gamma[A]&=&-\frac{1}{2}\lim_{C\to\infty}
\int^\infty_0\frac{dT}{T}\exp[-CT(\frac{D}{2}-1)]
\int_P{\cal D}x\
\frac{1}{2}(\int_A-\int_P){{\D}}\psi{\cal D}\bar\psi
\nonumber\\
&&\times 
{{\Tr}}{\PP}
\exp\biggl[-\int^T_0
d\tau\left\{\frac{1}{4}\dot x^2+ig A_\mu\dot x_\mu
+\bar\psi_\mu[(\partial_\tau-C)
\delta_{\mu\nu}-2ig F_{\mu\nu}]\psi_\nu\right\}
\biggr]
\nonumber\\
\ear
Several comments are in order here. The factor $\exp[-CTD/2]$ in
(\ref{3.27}) is due to the difference between the Weyl and
the anti-Wick-ordered Hamiltonian. It is crucial for obtaining
a finite result in the limit $C\to\infty$. In fact, for $D=4$ it
converts the prefactor $\exp[+CT]$ to a decaying exponential
$\exp[-CT]$ 
\footnote{In ref. \cite{str1} the reordering factor
was not taken into account and the change of the sign in $D=4$ was
 attributed to a difference between Minkowski and
Euclidean spacetime which is not correct in our opinion.}.
From the point of view of the worldline
fermions, $C$ plays the role of a mass.  Their free Green's
function ${\G}^C(\tau_2-\tau_1)\delta_{\mu\nu}$ with
\be\label{3.28}
{\G}^C(\tau_2-\tau_1)\equiv<\tau_2|(\partial_\tau-C)^{-1}|\tau_1>\ee
reads for periodic and antiperiodic boundary conditions, respectively,
\bear\label{3.29}
&&{\G}^C_P(\tau)=-[\Theta(-\tau)+\Theta(\tau)e^{-CT}]
\frac{e^{C\tau}}{1-e^{-CT}}\nonumber\\
&&{\G}^C_A(\tau)=-[\Theta(-\tau)-\Theta(\tau)e^{-CT}]
\frac{e^{C\tau}}{1+e^{-CT}}\ear
We observe that for $C\to\infty$ there is an increasingly
strong asymmetry between the forward and backward propagation
in the proper time. Further details of the Green's functions
(\ref{3.29}) can be found in appendix B.

We mention in passing that there exists another simple method
for the projection on the one-form sector. We can insert a
Kronecker-delta $\delta_{1\widehat F}$ into the partition function
(\ref{3.19}) and exponentiate  it in terms of a parameter
integral over an angular variable $\alpha$:
\be\label{3.29a}
{{\Tr}}_1\left[ e^{-T\widehat H_0}\right]=\frac{T}{2\pi}\int^{2\pi/T}_0
d\alpha \ e^{i\alpha T}\  {{\Tr}}\left[e^{-i\alpha T\widehat F}
e^{-T\widehat H_0}\right].\ee
The r.h.s.  of (\ref{3.29a}) can be represented by a path integral
 which, for the fermions,  involves antiperiodic boundary conditions
only.  The corresponding action is similar to the one used above but
with  $C$ replaced by $i\alpha$. Instead of the limit $C\to\infty$
one has to perform the $\alpha$-integration now. The computational
effort is essentially the same in both cases.

The representation (\ref{3.27}) of the effective action does
not coincide with the one used by Strassler \cite{str1}.
While he uses the same kinetic term in the fermionic worldline
Lagrangian, he modifies the interaction term according to
\be\label{3.30}
\bar\psi_\mu F_{\mu\nu}\psi_\nu\to
\half\chi_\mu
F_{\mu\nu}\chi_\nu\equiv\bar\psi_\mu
F_{\mu\nu}\psi_\nu+\frac{1}{2} F_{\mu\nu}
(\psi_\mu\psi_\nu+\bar\psi_\mu\bar\psi_\nu)\ee
where
\be\label{3.31}
\chi_\mu(\tau)\equiv\psi_\mu(\tau)+\bar\psi_\mu(\tau)\ee
As a consequence, the modified Hamiltonian $\widehat H_0$ contains terms
which change the fermion number by 2 units. This means that, during
the proper time evolution, the 1-form representing the gauge
boson can evolve into a 3-form. However, in the limit
$C\to\infty$ the
substitution (\ref{3.30}) causes no problems. The reason is that for
$C\to\infty$ the energy gap between 1- and 3-forms becomes infinite,
and therefore a 1-form at $\tau=0$ will remain a 1-form for
alle $\tau>0$. In the modified formalism,  Wick contractions of the
interaction term (\ref{3.3}) involve the 2-point 
function of $\chi$, i.e.,
${\G}^\chi(\tau)\equiv{\G}^C(\tau)-{\G}^C(-\tau)$. From
(\ref{3.29}) we obtain explicitly
\bear\label{3.32}
{\G}^\chi_P(\tau)&=&{\rm sign}(\tau)\frac{\sinh[C(\frac{T}{2}
-|\tau|)]}{\sinh[CT/2]}\nonumber\\
{\G}^\chi_A(\tau)&=&{\rm sign}(\tau)\frac{\cosh[C(\frac{T}{2}
-|\tau|)]}{\cosh[CT/2]}\ear
These Green's functions do not coincide with the ones given by
Strassler \cite{str1}; however,  they become effectively equivalent
in the limit $C\to\infty$. The substitution (\ref{3.30}) is
motivated by the algebraic simplification which it entails in
perturbation theory where one inserts a sum of plane
waves for $A^a_\mu(x)$. We shall not do this in the present
paper but rather calculate the path integral exactly for
a covariantly constant field strength. In this case the
representation (\ref{3.27}) is more convenient than the one
advocated by Strassler.

An important building block for  higher-loop calculations is the
gluon propagator in an external Yang-Mills field. It is given by
the proper-time integral of the evolution kernel (\ref{3.13}). The
latter can be represented by the following path integral with open
boundary conditions:
\be\label{3.33}
K(x_2,\psi_2,T|x_1,\psi_1,0)
= \int^{x(T)=x_2}_{x(0)=x_1}{\D} x(\tau) \int^{\psi(T)=\psi_2}_
{\psi(0)=\psi_1}{\D}\psi(\tau)\int{\D}\bar\psi(\tau) e^{-\int^T_0
d\tau L}.\ee
The Lagrangian $L$ is given by (\ref{3.16}) with $M_{\mu\nu}=-2ig
F_{\mu\nu}$ because we do not need the $C$-term in the case of
open boundary conditions. The reason is that the Hamiltonian
(\ref{3.8}) preserves the fermion number. Therefore, the kernel
$K$ will map a one-form wave function of the type (\ref{3.11})
onto another one-form. In fact, in the one-form sector, $K$ is
represented by a Lorentz matrix
\be\label{3.34}
K_{\mu\nu}(x_2,T|x_1,0)=\langle x_2,\mu|e^{-T{\D}}
|x_1,\nu\rangle\ee
which is related to the gluon propagator by
\be\label{3.35}
\langle x_2,\mu|{\D}^{-1}|x_1,\nu\rangle=\int^\infty_0
dT \ K_{\mu\nu}(x_2,T|x_1,0).\ee
(The color indices are kept implicit.) It is easy to
 express the bosonic matrix $K_{\mu\nu}$ in terms of
the kernel $K$ with fermionic arguments. If one writes
\be\label{3.36}
\Phi(x_2,\psi_2,T)=\int d^D x_1 d^D\psi_1 \ K(x_2,\psi_2,T|
x_1,\psi_1,0)\ \Phi(x_1,\psi_1,0)\ee
and inserts a wave function of the type (\ref{3.11}) at both
$\tau=0$ and $\tau=T$ one finds that
\be\label{3.37}
K_{\mu\nu}(x_2,T|x_1,0)=\frac{\partial}{\partial \psi_{2\mu}}
\int d^D\psi_1 \ K(x_2,\psi_2,T|x_1,\psi_1,0)\ \psi_{1\nu}.\ee
By combining eq. (\ref{3.35}) with eqs. (\ref{3.37}) and (\ref{3.33})
we obtain the desired path integral representation of the gluon
propagator.

In order to get a better understanding of this representation,
 let us assume
that the background $A_\mu$ is either abelian or quasi-abelian, and
that no path-ordering must be observed therefore. We may  then rewrite
(\ref{3.33}) according to

\bear\label{3.38}
K(x_2,\psi_2,T|x_1,\psi_1,0)
&=&\int^{x(T)=x_2}_{x(0)=x_1}{\D} x(\tau)\  K_F(\psi_2,T|
\psi_1,0)\nonumber\\
&&\times 
\exp\left[-\int^T_0 d\tau\left\{\frac{1}{4}
\dot x^2+ig A_\mu \dot x_\mu\right\}\right].\ear
Here the fermionic integral
\bear\label{3.39}
K_F(\psi_2,T|\psi_1,0)&=&\int^{\psi(T)=\psi_2}_{\psi(0)=\psi_1}
{\D}\psi(\tau)\int{\D}\bar\psi(\tau)\nonumber\\
&&\times\exp \left[-\int^T_0 d\tau\bar\psi_\mu
(\partial_\tau\delta_{\mu\nu}-2ig F_{\mu\nu})\psi_\nu\right]\ear
is a functional of the bosonic trajectory $x_\mu(\tau)$. It can be
evaluated exactly \cite{bag,cm,met} and has a remarkably simple
structure:
\be\label{3.40}
K_F(\psi_2,T|\psi_1,0)=\delta(\psi_2-{\cal S}(T)\psi_1).\ee
In (\ref{3.40}) we introduced
\be\label{3.41}
{\cal S}_{\mu\nu}(T)={\cal P}\exp\left[2ig\int^T_0
d\tau F(x(\tau))\right]_{\mu\nu}\ee
where the path-ordering is necessary because of the Lorentz matrix
structure. Alternatively, eq. (\ref{3.40}) can be established by noting
that $K_F$ solves the Schr\"odinger equation corresponding to
(\ref{3.39}):
\be\label{3.42}
\left[\frac{\partial}{\partial T}-2ig \psi_\mu F_{\mu\nu}
\frac{\partial}{\partial\psi_\nu}\right] K_F(\psi,T|
\psi_1,0)=0.\ee
Because ${\cal S}^T={\cal S}^{-1}$, it follows that
\be\label{3.43}
\frac{\partial}{\partial\psi_{2\mu}}\int d^D\psi_1
K_F(\psi_2,T|\psi_1,0)\psi_{1\nu}={\cal S}_{\mu\nu}(T)\ee
and therefore
\bear\label{3.44}
K_{\mu\nu}(x_2,T|x_1,0)&=&\int^{x(T)=x_2}_{x(0)=x_1}{\D} x(\tau)
\ {\cal P}\exp\left[ 2ig\int^T_0 d\tau F(x(\tau))\right]_{\mu\nu}
\nonumber\\
&&\times\exp\left[-\int^T_0d\tau\left\{\frac{1}{4}\dot x^2
+ig A_\mu\dot x_\mu\right\}\right].\ear
Later on we shall use the evolution kernel in the form
(\ref{3.44}). Clearly we could have written down this
representation without going through the original path integral
(\ref{3.33}). However, in multiloop calculations it will be
advantageous if the gluon loops and the corresponding propagators
 are represented in a coherent framework. In fact, if we recall
that $\bar\psi_\mu$ amounts to the derivative $\partial/\partial
\psi_\mu$ in the Schr\"odinger picture, the evolution kernel, for
any background, may be rewritten in the following very elegant form:
\bear\label{3.45}
K_{\mu\nu}(x_2,T|x_1,0)&=&
\int^{x(T)=x_2}_{x(0)=x_1}{\D}x(\tau) \int {\D}  
\psi(\tau)\int{\D}\bar\psi(\tau)\nonumber\\
&&\times
\delta(\psi
(T))\ \bar\psi_\mu(T)\psi_\nu(0)\ e^{-\int^T_0d\tau L}.\ear
In eq. (\ref{3.45}), $\psi(0)$ and $\psi(T)$ are integrated
 independently.

Now we evaluate the path integral (\ref{3.27}) for the case that
the background has a covariantly constant field strength. We assume
that the gauge field has the form
\be\label{4.101}
A^a_\mu(x)=n^a{\sf A}_\mu(x),\quad n^an^a=1\ee
where $n^a$ is a constant unit vector in color space. The
 associated field strength $F^a_{\mu\nu}=n^a{\sf F}
_{\mu\nu}$ with ${\sf F}_{\mu\nu}=\partial_\mu{\sf A}_\nu
-\partial_\nu{\sf A}_\mu$ satisfies $D^{ab}_\alpha F^b_{\mu\nu}
=0$.  Both $A_\mu$ and $F_{\mu\nu}$ enter eq. (\ref{3.27}) for
the gauge boson loop as matrices in the adjoint representation.
Hence
the only nontrivial color matrix which enters the path integral
is $n^aT^a$. If we denote the eigenvalues of this matrix by $\nu
_l, l=1,2,...$, and evaluate the color trace in the diagonal
basis, the integral of the path integrand assumes the form

\be\label{4.102}
{{\Tr}} {\PP}I(gA_\mu)=\sum_l I(g\nu_l{\sf A}_\mu)\ee
The path ordering has no effect here. We observe that we
are effectively dealing with a set of abelian theories whose
gauge coupling is given by $g\nu_l$.

While the condition $D_\alpha^{ab}F_{\mu\nu}^b=0$ does not
necessarily imply that ${\sf F}_{\mu\nu}$ is constant,
we further assume that ${\sf F}_{\mu\nu}$ is an
$x$-independent matrix and that the gauge field is of the
form
\be\label{4.103}
{\sf A}_\mu(x)=\frac{1}{2}x_\nu {\sf F}_{\nu\mu}\ee
In the following we keep the diagonalization of the color
matrix implicit and continue to use the notation ${{\Tr}}
I(gA_\mu)$ rather than $\sum_lI(g\nu_l{\sf A}_\mu)$.
We keep in mind, however, that $A_\mu$ and $F_{\mu\nu}$ may be
treated as pure numbers as far as their color structure is
concerned.

For the field (\ref{4.101}), (\ref{4.103}) all path
integrals in (\ref{3.27}) are Gaussian. We separate the constant
mode from the $x_\mu$ integration as before and obtain
\be\label{4.104}
\Gamma[A]=-\frac{1}{2}\int d^Dx_0{\rm tr}\int^\infty_0\frac{
dT}{T}{\Det_P'}[-\partial^2_\tau\delta_{\mu\nu}+2ig F_{\mu\nu}
\partial_\tau]^{-\frac{1}{2}}
\lim_{C\to\infty}Y(C)\ee
with
\bear\label{4.105}
Y(C)&=&\frac{1}{2}\exp\Bigl
[-CT(\frac{D}{2}-1)\Bigr]\biggl\lbrace{\Det_A}[(\partial_\tau-C)
\delta_{\mu\nu}-2igF_{\mu\nu}\bigr]\nonumber\\
&&-{\Det_P}\bigl[(\partial_\tau-C)\delta_{\mu\nu}-2igF_{\mu\nu}
\bigr]\biggr\rbrace
\ear
We denote the real eigenvalues of $iF_{\mu\nu}$ by
$f^{(\alpha)},\alpha=1,...,D$, and we use the formulas in appendix
C in order to express $Y(C)$ in terms of these eigenvalues:
\bear\label{4.106}
&&Y(C)=\frac{1}{2}\exp[-CT(\frac{D}{2}-1)]\nonumber\\
&&\times\Biggl\{{{{\Det}}}_A[(\partial_\tau-C)\delta_{\mu\nu}]
\prod_{\alpha=1}^D\frac{\cosh[\frac{T}{2}(C+2gf^{(\alpha)})]}
{\cosh[CT/2]}\nonumber\\
&&-{{\Det}}_P[(\partial_\tau-C)\delta_{\mu\nu}]
\prod^D_{\alpha=1}\frac{\sinh[\frac{T}{2}(C+2g f^{(\alpha)})]}
{\sinh[CT/2]}\Biggr\}\ear
Note that in the case of the fermionic integration with periodic
boundary conditions the zero mode of $\partial_\tau$ was {\it not}
excluded from the determinant and that eq. (\ref{C.6}) applies
therefore. In (\ref{4.106}) we have factored out the free
determinants because the method of appendix C can yield only ratios
of determinants. The normalization factors
\be\label{4.107}
\zeta_{A,P}\equiv \Det{}_{A,P}[(\partial_\tau-C)\delta_{\mu\nu}]
=\left(\Det{}_{A,P}[\partial_\tau-C]\right)^D\ee
are most easily determined by recalling their operatorial
interpretation as the partition function and the Witten
index of a free Fermi oscillator, respectively:
\bear\label{4.108}
&&{{\Det}}_A[\partial_\tau-C]={{\Tr}}[e^{-CT\widehat F_W}]\nonumber\\
&&{{\Det}}_P[\partial_\tau-C]
={{\Tr}}[(-1)^{\widehat F}e^{-CT\widehat F_W}]\ear
Here $\widehat F_W
=(\widehat\psi\widehat{\bar\psi}-\widehat{\bar\psi}\widehat\psi)/2
=\widehat F-1/2$ is the Weyl-ordered fermion number operator (for $D=1$)
with eigenvalues -1/2 and +1/2. As a consequence,
\bear\label{4.109}
\zeta_A&=&(2\cosh[CT/2])^D\nonumber\\
\zeta_P&=&(2\sinh[CT/2])^D\ear
Taking advantage of $\Sigma_\alpha f^{(\alpha)}=0$, we can
rewrite (\ref{4.106}) as
\be\label{4.110}
Y(C)=\frac{1}{2}\eta^{-1}\left[\prod_\alpha(1+\eta q_\alpha)-\prod_
\alpha(1-\eta q_\alpha)\right]\ee
with $\eta\equiv\exp(-CT)$ and $q_\alpha\equiv\exp[-2g Tf^{(\alpha)}]$.
In the limit $C\to\infty$, i.e.,  $\eta\to 0$, the leading $O(1)$
terms cancel among the products in (\ref{4.110}), and one obtains
\bear\label{4.111}
\lim_{C\to\infty}Y(C)&=&\sum^D_{\alpha=1}\exp[-2g T f^{(\alpha)}]
\nonumber\\
&=&{\rm tr}_L\cos[2g T F]\ear
Here we exploited that if $f^{(\alpha)}$ is an eigenvalue, so
is $-f^{(\alpha)}$. (${\rm tr}_L$ denotes the trace with respect
to the Lorentz indices.)

In complete analogy with the scalar 
case, eq.(~\ref{scaldetcomp}), the bosonic
determinant in (\ref{4.104}) gives rise to a factor of
\be\label{4.112}
(4\pi T)^{-D/2}\exp\left[-\frac{1}{2}{{\rm tr}}_L\ln
\frac{\sin(gTF)}{(gTF)}\right]\ee
Hence our final result for the gauge boson loop becomes
\bear\label{4.113}
\Gamma[A]&=&-\frac{1}{2}\int d^Dx_0\ (4\pi)^{-D/2}\int^\infty
_0{dT\over T} T^{-D/2}\nonumber\\
&&\quad\times{{\rm tr}}\exp\left[-\frac{1}{2}{{\tr}}_L\ln\frac{\sin(g
TF)}{(gTF)}\right]\nonumber\\
&&\quad\times {{\rm tr}}_L\cos(2gTF)\ear
Eq. (\ref{4.113}) coincides with the result which was
found with the help of the traditional techniques 
\cite{shore,bat,drgcd,bvw}.

By a computation similar to the previous one, but with open boundary
conditions for the path integral, one can find the gluon propagator
in the background (\ref{4.101}), (\ref{4.103}). It is given by
(\ref{3.35}) with
\be\label{4.114}
K_{\mu\nu}(x_2,T|x_1,0)=\exp (2ig TF)_{\mu\nu}
\int^{x(T)=x_2}_{x(0)=x_1}{\D} x(\tau) e^{-S[x(\cdot)]}.\ee
The path integral which remains to be evaluated
ist just the one for the
scalar propagator. Now
in the action functional
\bear\label{4.115}
S[x(\cdot)]&=&\frac{1}{4}\int^T_0 d\tau \ x_\mu\left\{
-\partial^2_\tau\delta_{\mu\nu}+2ig F_{\mu\nu} \partial_\tau
\right\} x_\nu\nonumber\\
&&+\frac{1}{4}\left\{ x_\mu(T)\dot x_\mu(T)-x_\mu(0)\dot x_\mu
(0)\right\}\ear
one has to be careful about the surface terms which  appear after the
integration by parts. For open paths they give rise to a non-zero
contribution in general. Since $S$ is quadratic in $x_\mu$, the
saddle point approximation of the path integral (\ref{4.114}) gives
the exact answer. We can solve it by expanding the integration
variable about the classical trajectory connecting $x_1$ and $x_2$:
\be\label{4.115a}
x_\mu(\tau)=x_\mu^{\rm class}(\tau)+y_\mu(\tau)\ee
Here $x^{\rm class}(\tau)$ obeys
\be\label{4.116}
\ddot{x}_\mu^{\rm class}=2ig F_{\mu\nu}\dot x_\nu^{\rm class}\ee
and it satisfies the boundary conditions $x^{\rm class}(0)=x_1$
and $x^{\rm class}(T)=x_2$. The fluctuation $y(\tau)$ satisfies
correspondingly $y(0)=0=y(T)$. Hence the fluctuation determinant
is almost the same as in the periodic case, the only difference
being that there is no zero-mode integration in the present
case:
\bear\label{4.117}
K_{\mu\nu}(x_2,T|x_1,0)&=&\exp(2ig TF)_{\mu\nu}\ \exp\left(-S
[x^{\rm class}]\right)\nonumber\\
&&\times {\Det{}^{\prime}_P}\left[-\partial^2_\tau \delta_{\mu\nu}+2ig
F_{\mu\nu}\partial_\tau\right]^{-1/2}.\ear
The classical trajectory is easily found:
\be\label{4.118}
x^{\rm class}(\tau)=x_1+\frac{\exp(2ig F\tau)-1}{\exp(2ig FT)-1}
(x_2-x_1).\ee
Its action is entirely due to the surface terms in (\ref{4.115}).
In Fock-Schwinger gauge centered at $x_1$
it reads
\be\label{4.119}
S[x^{\rm class}]=\frac{1}{4}(x_2-x_1) gF\cot (gTF) (x_2-x_1).\ee
Putting everything together we obtain the final result for the
propagation kernel:
\bear\label{4.120}
K_{\mu\nu}(x_2,T|x_1,0)
&=&(4\pi T)^{-D/2}\exp[2ig TF]_{\mu\nu}\nonumber\\
&&\times\exp\left[-\frac{1}{4}(x_2-x_1) 
gF\cot(gTF)(x_2-x_1)\right]
\nonumber\\
&&\times\exp\left[-\frac{1}{2}{\rm tr}_L
\ln\frac{\sin(gTF)}{(gTF)}
\right].\ear

\vfill\eject

\vskip15pt
{\bf 4. A Multiloop Generalization for Quantum Electrodynamics}
\renewcommand{\theequation}{4.\arabic{equation}}
\setcounter{equation}{0}
\vskip10pt

We return to the case of quantum electrodynamics,
and proceed to the multiloop generalization of
the formalism developed in chapter 2.
This generalization is constructed in
strict analogy to the case without an external field
~\cite{ss3}, and we refer the reader to that publication
for some of the details.

We first consider scalar electrodynamics at the two-loop level,
i.e. a scalar loop with an internal photon correction. 
A photon insertion in
the worldloop may, in Feynman gauge, be represented in terms of
the following current-current interaction term 
(see e.g. ~\cite{fried,makmig,dorn})
inserted into
the one-loop path integral,

\begin{equation}
-{e^2\over 2}
{{\Gamma (\lambda )}\over {4{\pi}^{\lambda +1}}}
\int_0^T d\tau_a \int_0^T d\tau_b
{{\dot x(\tau_a)\cdot\dot x(\tau_b)}\over
{\Bigl({[x(\tau_a) - x(\tau_b)]}^2\Bigr)}^{\lambda}}\quad 
\label{cci}
\end{equation}

\noindent
($\lambda = {D\over2} -1$).
The denominator of this term is again
written in the proper-time
representation,

\begin{equation}
{\Gamma(\lambda )\over {4{\pi}^{\lambda +1}}
{\Bigl({[x(\tau_a) - x(\tau_b)]}^2\Bigr)}^{\lambda}}
=\int_0^{\infty} d\bar T 
{(4\pi \bar T)}^{-{D\over 2}}
{\rm exp}\Biggl
[-{{\Bigl(x(\tau_a)-x(\tau_b)\Bigr)}^2\over 4\bar T}
\Biggr ]\, .
\label{intprop}
\end{equation}
It appears then as yet another correction term to the
free part of the worldline Lagrangian for the scalar
loop path integral. It is convenient to rewrite this
term in the form

\begin{equation}
{(x(\tau_a)-x(\tau_b))}^2
= 
\int_0^T d\tau_1 \int_0^T d\tau_2
\,x(\tau_1)B_{ab}(\tau_1,\tau_2)x(\tau_2)
\quad ,
\label{useB}
\end{equation}
\noindent
with
\begin{equation}
B_{ab}(\tau_1,\tau_2) = 
\Bigl[\delta(\tau_1 - \tau_a)-
\delta(\tau_1-\tau_b)\Bigr]
\Bigl[\delta(\tau_2 - \tau_a)-
\delta(\tau_2-\tau_b)\Bigr]
\quad .
\label{defB}
\end{equation}
\noindent
This allows us to write the new total 
bosonic kinetic operator as
\begin{equation}
\partial^2
-2ieF\partial
-{B\over\bar T}\quad.
\label{kinFB}
\end{equation}
\noindent
To find the inverse of this operator, we write it
as a geometric series,

\begin{equation}
{\Bigl(\partial^2 -2ieF\partial
-{B\over\bar T} \Bigl)}^{-1}
= {\Bigl(\partial^2 -2ieF\partial\Bigr)}^{-1}
+
{\Bigl(\partial^2 -2ieF\partial\Bigr)}^{-1}
{B_{ab}\over \bar T}
{\Bigl(\partial^2 -2ieF\partial\Bigr)}^{-1}
+\cdots,
\label{sumseries}
\end{equation}
\no
which can be easily summed to yield the following
Green's function:

\begin{equation}
{\cal G}_B^{(1)}(\tau_1,\tau_2)=
{\cal G}_B(\tau_1,\tau_2) + {1\over 2}
{{[{\cal G}_B(\tau_1,\tau_a)-{\cal G}_B(\tau_1,\tau_b)]
[{\cal G}_B(\tau_a,\tau_2)-{\cal G}_B(\tau_b,\tau_2)]}
\over
{{\bar T} -{1\over 2}
{\cal C}_{ab}}}\,
\label{calG(1)} 
\end{equation}
\noindent
with the definition

\begin{eqnarray}
{\cal C}_{ab}&\equiv& 
{\cal G}_B(\tau_a,\tau_a)
-{\cal G}_B(\tau_a,\tau_b)
-{\cal G}_B(\tau_b,\tau_a)
+{\cal G}_B(\tau_b,\tau_b)
\nonumber\\
&=&
{\cos (eFT)-\cos (eFT\dot G_{Bab})
\over (eFT)\sin (eFT)}
\; .\nonumber\\
\label{defCab}
\end{eqnarray}
\noindent
Note that this is almost identical with what one would
obtain from the ordinary bosonic
two-loop Green's function $G_B^{(1)}$ ~\cite{ss2}
by simply replacing all $G_{Bij}$'s appearing there
by the corresponding
${\cal G}_{Bij}$'s. The more complicated structure of the
denominator is due to the fact that
the ${\cal G}_{Bij}$'s are not symmetric anymore,
rather we have
${\cal G}_{Bij}={\cal G}_{Bji}^T$, and moreover
have nonvanishing coincidence limits.
The denominator is now in general a nontrivial
Lorentz matrix, and must be interpreted as
a matrix inverse (of course, all matrices
appearing here commute with each other).

For the free Gaussian path integral,
it is again a simple application of the 
$ln\,det = tr\,log$ -- identity to calculate

\begin{eqnarray}
{\rm Det'}_P
\biggl[ -{\partial}^2+2ieF{\partial}+{B\over\bar T}\biggr]
&=&
{\rm Det'}_P
\Bigl[-{\partial}^2\Bigr]
{\rm Det'}_P
\biggl[{\bf I}-2ieF{\partial}^{-1}
\biggr]
\nonumber\\
&&\times
{\rm Det'}_P
\biggl[
{\bf I}-{B\over\bar T}
{\Bigl(
{\partial}^2-2ieF{\partial}\Bigr)
}^{-1}
\biggr]
\nonumber\\
&=&
{\lbrack 4\pi T\rbrack}^D
{\rm det}\biggl[{\sin(eFT)\over{eFT}}\biggr]
{\rm det}
\biggl[{{\bf I}
-{1\over {2\bar T}}
{\cal C}_{ab}
}\biggr]
\quad .\nonumber\\
\label{FBbosnorm}
\end{eqnarray} 
\noindent
The generalization from one-loop to two-loop
photon amplitude calculations in scalar QED thus
requires no changes of the formalism itself,
but only of the Green's functions used, and of
the global determinant factor. Of course,
in the end three more parameter integrations have
to be performed.

As in the case without a background field ~\cite{ss3},
the whole procedure goes through essentially unchanged
for the fermion loop, if the superfield 
formalism is used.
As a consequence,
the supersymmetrization property carries over
to the two-loop level, leading to a
close relationship between the parameter
integrals for the same amplitude 
calculated for the scalar and for the
fermion loop: They differ only
by a replacement of all ${\cal G}_B$'s
by $\hat{\cal G}$'s, 
by the additional $\theta$ -- 
integrations, and by the one-loop Grassmann path integral
factor eq.(~\ref{grassfermcomp}).

The generalization to an arbitrary fixed number of photon insertions
is straightforward, and leads to
formulas for the generalized (super-) Green's functions 
and (super-) determinants identical with the ones given in
\cite{ss2,ss3} up to a replacement of all
$G_B$'s ($\hat G$'s) by ${\cal G}_B$'s
($\hat{\cal G}$'s). The only point to be mentioned here
is that care must now be taken in writing the indices of
the ${\cal G}_{Bij}$'s appearing. For instance, the 
bosonic three-loop
Green's function must be written

\begin{eqnarray}
{\cal G}^{(2)}_B(\tau_1,\tau_2)
=
{\cal G}_B(\tau_1,\tau_2)
\!
+\frac{1}{2}
\sum_{k,l=1}^2
\Bigl[
{\cal G}_B(\tau_1,\tau_{a_k})
-{\cal G}_B(\tau_1,\tau_{b_k})
\Bigr]
A_{kl}^{(2)}
\Bigl[
{\cal G}_B(\tau_{a_l},\tau_2)
-{\cal G}_B(\tau_{b_l},\tau_2)
\Bigr].
\nonumber\\
\label{3loopGreen's}
\end{eqnarray}
\noindent
The matrix $A$ appearing here
is the inverse of the matrix

\begin{equation}
\left(
\begin{array}{*{2}{c}}
T_1-{1\over 2}
\Bigl(\G_{Ba_1a_1}-\G_{Ba_1b_1}-\G_{Bb_1a_1}+\G_{Bb_1b_1}\Bigr)
&
-\half\Bigl(\G_{Ba_1a_2}-\G_{Ba_1b_2}-\G_{Bb_1a_2}+\G_{Bb_1b_2}\Bigr)
\\
-\half
\Bigl(\G_{Ba_2a_1}-\G_{Ba_2b_1}-\G_{Bb_2a_1}+\G_{Bb_2b_1}\Bigr)
&
T_2-\half
\Bigl(\G_{Ba_2a_2}-\G_{Ba_2b_2}-\G_{Bb_2a_2}+\G_{Bb_2b_2}\Bigr)\\
\end{array}
\right)\\
\label{3loopA}
\end{equation}
\noindent
$T_1,T_2$ denote the proper-time lengths
of the two inserted propagators.

The discussion of the general case of an arbitrary number of
scalar (spinor) loops interconnected by photon propagators
requires no new concepts, and 
will be deferred to a forthcoming review article
~\cite{ss4}.

While this multiloop construction is done most simply using
the Feynman gauge for the propagator insertions, other
gauges can be implemented as well (the gauge freedom has also
been discussed in ~\cite{dss}).
In an arbitrary covariant gauge, the photon insertion term
eq.(~\ref{cci}) would read

\begin{eqnarray}
-{e^2\over 2}
{1\over {4{\pi}^{{D\over2}}}}
\int_0^T d\tau_a \int_0^T d\tau_b
\Biggl\lbrace
{{1+\alpha}\over 2}\Gamma\Bigl({D\over 2}-1\Bigr)
{{\dot x_a\cdot\dot x_b\over
{\Bigl[{(x_a - x_b)}^2\Bigr]}^{{D\over 2}-1}}\quad
}\nonumber\\
+(1-\alpha)\Gamma\Bigl({D\over 2}\Bigr)
{\dot x_a\cdot(x_a-x_b)
(x_a-x_b)\cdot\dot x_b\over
{\Bigl[{(x_a - x_b)}^2\Bigr]}^{D\over 2}\quad
}
\Biggr\rbrace
\label{ccigengauge}
\end{eqnarray}
\noindent
Here $\alpha = 1$ corresponds to Feynman gauge,
$\alpha = 0$ to Landau gauge.
The integrand may also be written as

\begin{equation}
\Gamma\Bigl({D\over 2}-1\Bigr)
{{\dot x_a\cdot\dot x_b\over
{\Bigl[{(x_a - x_b)}^2\Bigr]}^{{D\over 2}-1}}\quad
}\nonumber\\
-{1-\alpha\over 4}
\Gamma\Bigl({{D\over 2}-2}\Bigr)
{\partial\over\partial\tau_a}
{\partial\over\partial\tau_b}
{\Bigl[{(x_a-x_b)}^2\Bigr]}
^{2-{D\over2}}
.
\label{ccigengaugetotdir}
\end{equation}
\noindent
This shows that on the worldline gauge transformations
correspond to the addition of total derivative
terms, which is another
fact familiar to string theorists 
(see e.g. ~\cite{sathiapalan}). This form of the
photon insertion is also the more practical one
for actual calculations.
Again it carries over to the fermion loop in the
superfield formalism {\it mutatis mutandis}.

\vskip15pt
\vfill\eject
{\bf 5. The Two-Loop Euler-Heisenberg Lagrangian
for Scalar QED}
\renewcommand{\theequation}{5.\arabic{equation}}
\setcounter{equation}{0}
\vskip10pt
We proceed to the simplest two-loop application
of this formalism, which is the two-loop generalization of
Schwinger's formula for the constant field effective Lagrangian
due to a scalar loop, eq.(~\ref{eulheiscal}).
According to the above, we may write the two-loop
correction to this effective Lagrangian in the form

\begin{eqnarray}
{\cal L}^{(2)}_{\rm scal}
[F]&=&
{(4\pi )}^{-D}
\Bigl(-{e^2\over 2}\Bigr)
\int_0^{\infty}{dT\over T}e^{-m^2T}T^{-{D\over 2}} 
\int_0^{\infty}d\bar T 
\int_0^T d\tau_a
\int_0^T d\tau_b
\nonumber\\
&\phantom{=}&\times
{\rm det}^{-{1\over 2}}
\biggl[{\sin(eFT)\over {eFT}}
\biggr]
{\rm det}^{-{1\over 2}}
\biggl[
\bar T 
-{1\over 2}
{\cal C}_{ab}
\biggr]
\langle
\dot y_a\cdot\dot y_b\rangle
\quad .\nonumber\\
\label{Gamma2scal}
\end{eqnarray}
\noindent
We have now a fourfold parameter integral, with
$T$ and $\bar T$ representing the scalar 
and photon proper-times, and $\tau_{a,b}$ the
endpoints of the photon insertion moving around
the scalar loop. The first determinant factor
is identical with the one-loop Euler-Heisenberg-Schwinger
integrand eq.(~\ref{scaldetcomp}), 
and represents the change of the free
path integral
determinant due to the external field;
the second one 
represents its change due to the photon insertion. 
A single Wick contraction needs to be
performed
on the ``left over'' numerator of the
photon insertion, using the modified worldline
Green's function eq.(~\ref{calG(1)}). This yields

\begin{equation}
\langle
\dot y_a\cdot\dot y_b\rangle
=
{\rm tr}
\biggl[
\ddot{\cal G}_{Bab}+{1\over 2}
{(\dot {\cal G}_{Baa}-\dot {\cal G}_{Bab})
(\dot {\cal G}_{Bab}-\dot {\cal G}_{Bbb})
\over
{\bar T -{1\over 2}{\cal C}_{ab}}}
\biggr]\; .
\label{wickscal}
\end{equation}
\noindent
Care must be taken again with coincidence limits, as the
derivatives should {\sl not} act on the variables
$\tau_a,\tau_b$ explicitly appearing in the
two-loop Green's function; again the correct rule 
in calculating
$\langle\dot y_a\dot y_b\rangle$ is to first differentiate
eq.(~\ref{calG(1)}) with respect to $\tau_1,\tau_2$,
and put $\tau_1=\tau_a,\tau_2=\tau_b$ afterwards.

After replacing the $\dot {\cal G}_{Bij}$'s 
and ${\cal C}_{ab}$
by the explicit expressions given in eqs.(~\ref{derivcalGB})
and eq.(~\ref{defCab}),
we have already a parameter integral representation for
the bare dimensionally regularized effective Lagrangian.

Alternatively one may, in the spirit of the 
original Bern-Kosower approach,
remove $\ddot {\cal G}_B$ by a
partial integration with respect to $\tau_a$ or $\tau_b$.
Using the formula

\begin{equation}
d\det(M)=\det(M){\rm tr}(dMM^{-1})
\label{ddet}\nonumber
\end{equation}
\noindent
and $\dot {\cal G}_{Bab}=-\dot{\cal G}_{Bba}^T$, 
one obtains the equivalent parameter integral

\begin{eqnarray}
{\cal L}^{(2)}_{\rm scal}[F]&=&
{(4\pi )}^{-D}
\Bigl(-{e^2\over 2}\Bigr)
\int_0^{\infty}{dT\over T}e^{-m^2T}T^{-{D\over 2}} 
\int_0^{\infty}d\bar T 
\int_0^T d\tau_a
\int_0^T d\tau_b
\nonumber\\
&\phantom{=}&\times
{\rm det}^{-{1\over 2}}
\biggl\lbrack{\sin(eFT)\over {eFT}}
\biggr\rbrack
{\rm det}^{-{1\over 2}}
\biggl[
\bar T -{1\over 2}
{\cal C}_{ab}
\biggr\rbrack\nonumber\\
&\phantom{=}&\times
{1\over 2}
\Biggl\lbrace
{\rm tr}\dot{\cal G}_{Bab}
{\rm tr}
\biggl\lbrack
{\dot{\cal G}_{Bab}
\over
{\bar T -{1\over 2}{\cal C}_{ab}}}
\biggr\rbrack
+{\rm tr}
\biggl[
{(\dot {\cal G}_{Baa}-\dot {\cal G}_{Bab})
(\dot {\cal G}_{Bab}-\dot {\cal G}_{Bbb})
\over
{\bar T -{1\over 2}{\cal C}_{ab}}}
\biggr]
\Biggr\rbrace\; .
\nonumber\\
\label{Gamma2scalpI}
\end{eqnarray}
\noindent

For the further evaluation 
and renormalization of this Lagrangian,
we will specialize the constant
field $F$ to a pure magnetic and to a pure electric
field in turns. To facilitate comparison with previous
calculations ~\cite{ritusscal,lebedev,ginzburg,ritusspin,ditreu}, we will 
moreover switch from
dimensional regularization to proper-time regularization.
This means that henceforth
we put $D=4$, and instead
introduce a proper-time UV cutoff $T_0$ 
later on.

We begin with a pure magnetic field.
The field is taken along the z-axis, so that
$F^{12}=B$,$F^{21}=-B$ are the only
non-vanishing components of
the field strength tensor. We also
introduce the abbreviations
$z=eBT$, and projection matrices

\begin{equation}
{\bf\hat F}\equiv
\left(
\begin{array}{*{4}{c}}
0&0&0&0\\
0&0&1&0\\
0&-1&0&0\\
0&0&0&0
\end{array}
\right),
{\bf I_{03}}\equiv
\left(
\begin{array}{*{4}{c}}
1&0&0&0\\
0&0&0&0\\
0&0&0&0\\
0&0&0&1
\end{array}
\right),
{\bf I_{12}}\equiv
\left(
\begin{array}{*{4}{c}}
0&0&0&0\\
0&1&0&0\\
0&0&1&0\\
0&0&0&0
\end{array}
\right).\nonumber\\
\label{defprojectors}
\end{equation}
\no
We may then rewrite the determinant factors
eqs.(2.34),(\protect\ref{Ffermnorm})
as

\bear
{\rm det}^{-{1\over 2}}
\biggl[{\sin(eFT)\over {eFT}}
\biggr]&=&
{z\over{\sinh(z)} }
\;\label{scalardetextB},\\ 
{\rm det}^{-{1\over 2}}
\biggl[{\tan(eFT)\over {eFT}}
\biggr]&=&
{z\over{\tanh(z)} }
\; .
\label{spinordetextB}
\ear\no
The Green's functions eq.(\ref{calGBGF}),(\ref{derivcalGB}) 
specialize to

\begin{eqnarray}
\bar{\cal G}_{B}(\tau_1,\tau_2) 
&=& G_{B12}{\bf I_{03}}
-{T\over 2}{\Bigl[\cosh(z\dot G_{B12})-\cosh(z)\Bigr]
\over z\sinh(z)}
{\bf I_{12}}\nonumber\\&&
+{T\over{2z}}\biggl({\sinh(z\dot G_{B12})\over\sinh(z)}
-\dot G_{B12}\biggr)i{\hat{\bf F}}\nonumber\\
\dot{\cal G}_{B}(\tau_1,\tau_2)
&=&\dot G_{B12}{\bf I_{03}}+{\sinh(z\dot G_{B12})\over\sinh(z)}
{\bf I_{12}}
-\biggl({\cosh(z\dot G_{B12})\over \sinh(z)}-{1\over z}
\biggr)i{\hat{\bf F}}\nonumber\\
\ddot{\cal G}_{B}(\tau_1,\tau_2)
&=& \ddot G_{B12}{\bf I_{03}}
+2\biggl(\delta_{12}-{z\cosh(z\dot G_{B12})\over T
\sinh(z)}\biggr){\bf I_{12}}
+2{z\sinh(z\dot G_{B12})\over T\sinh(z)}i{\hat{\bf F}}\nonumber\\
{\cal G}_{F}(\tau_1,\tau_2) &=&G_{F12}{\bf I_{03}}
+G_{F12}{\cosh (z\dot G_{B12})\over \cosh (z)}{\bf I_{12}}
-G_{F12}{{\sinh (z\dot G_{B12})}\over{\cosh (z)}}i{\hat{\bf F}}
\nonumber\\
\label{GB(F)pureB}
\end{eqnarray}
\noindent
In writing ${\cal G}_B$ we have already subtracted its
coincidence limit, which is indicated by the ``bar''.
\no
$C_{ab}$ simplifies to 

\begin{equation}
{\cal C}_{ab}=-2G_{Bab}{\bf I_{03}}
-2G_{Bab}^z
{\bf I_{12}}\; ,
\label{simplifyCab}
\end{equation}
\noindent
where we have defined

\begin{equation}
G_{Bab}^z\equiv {T\over 2}
{\Bigl[\cosh (z)-\cosh(z\dot G_{ab})\Bigr]\over
z\sinh (z)}
=G_{Bab}-{1\over 3T}G_{Bab}^2z^2+O(z^4)
\label{abbrzGz}
\end{equation}
\noindent
We will also use the derivative of this expression,

\be
\dot G_{Bab}^z={\sinh(z\dot G_{Bab})\over\sinh(z)}
\label{abbrzdotGz}
\ee\no
Similarly we can rewrite

\begin{eqnarray}
{\rm det}^{-{1\over 2}}
\biggl[{\sin(eFT)\over {eFT}}{\bigl(\bar T 
-{1\over 2}
{\cal C}_{ab}
\bigr )}
\biggr]
&=&
{z\over\sinh(z)}
\gamma\gamma^z\nonumber\\
{\rm tr}\Bigl[\ddot{\cal G}_{Bab}\Bigr]
&=&
8\delta_{ab}-4-4{z\cosh(z\dot G_{Bab})\over\sinh(z)}
\nonumber\\
{1\over 2}{\rm tr}\dot{\cal G}_{Bab}
{\rm tr}\biggl[
{\dot{\cal G}_{Bab}\over
{\bar T -{1\over 2}{\cal C}_{ab}}}
\biggr]
&=& 2
\biggl[
\dot G_{Bab}+{\sinh(z\dot G_{Bab})\over\sinh(z)}
\biggr]
\biggl[
\dot G_{Bab}\gamma
+{\sinh(z\dot G_{Bab})\over \sinh(z)}
\gamma^z
\biggr]
\nonumber\\
{1\over 2}{\rm tr}
\biggl[
{(\dot {\cal G}_{aa}-\dot {\cal G}_{ab})
(\dot {\cal G}_{ab}-\dot {\cal G}_{bb})
\over
{\bar T -{1\over 2}{\cal C}_{ab}}}
\biggr]
&=&
-\gamma^z
{\sinh^2(z\dot G_{Bab})
+{\Bigl[\cosh(z\dot G_{Bab})-\cosh(z)\Bigr]}^2
\over
\sinh^2(z)}
\nonumber\\
&&-\dot G_{Bab}^2\gamma\nonumber\\
\label{simplifyremainder}
\end{eqnarray}
\noindent
with the abbreviations

\begin{eqnarray}
\gamma&\equiv&{(\bar T +G_{Bab})}^{-1},\nonumber\\
\gamma^z&\equiv&{(\bar T +G_{Bab}^z)}^{-1}.\nonumber\\
\label{defgamma}\nonumber
\end{eqnarray}
\noindent
We rescale to the unit circle,
$\tau_{a,b}=Tu_{a,b}$,
and use translation invariance in $\tau$ to
set $\tau_b=0$. We have then

\begin{eqnarray}
G_B(\tau_a,\tau_b)&=&TG_B(u_a,u_b)=T(u_a-u_a^2),\nonumber\\
\dot G_B(\tau_a,\tau_b)&=&\dot G_B(u_a,u_b)
=1-2u_a
\; .\nonumber\\
\label{scaleG}\nonumber
\end{eqnarray}
\noindent
After performance of the $\bar T$ -- integration,
which is finite and elementary,
eq.(~\ref{Gamma2scalpI})
turns into 

\begin{equation}
{\cal L}^{(2)}_{\rm scal}
[B]=-
{(4\pi )}^{-4}
{e^2\over 2}
\int_{0}^{\infty}{dT\over T^3}e^{-m^2T}
{z\over \sinh(z)}
\int_0^1 du_a
\,A(z,u_a)
\; ,
\label{Gamma2scalB}
\end{equation}
\noindent
with 

\begin{eqnarray}
A&=&
\Biggl\lbrace
A_1
{\ln ({G_{Bab}/ G_{Bab}^z})
\over{(G_{Bab}-G_{Bab}^z)}^2}
+{A_2\over
(G^z_{Bab})(G_{Bab}-G^z_{Bab})}
+{A_3\over
(G_{Bab})(G_{Bab}-G^z_{Bab})}
\Biggr\rbrace\quad ,\nonumber\\
A_1&=&
4
\Bigl[
G_{Bab}^z z\coth(z)-G_{Bab}
\Bigr]
\quad , \nonumber\\
A_2&=&1+
2\dot G_{Bab}
\dot G_{Bab}^z
-4G^z_{Bab}z\coth(z)
\quad ,\nonumber\\
A_3&=&
-\dot G_{Bab}^2-
2\dot G_{Bab}
\dot G_{Bab}^z
\; . 
\nonumber\\
\label{A1A2A3}
\end{eqnarray}
\noindent
$G_{Bab}^z$ is now given by eq.(5.10) with $T=1$.
Here and in the following we often 
use the identity
$\dot G_{Bab}^2=1-4G_{Bab}$ to
eliminate $\dot G_{Bab}$ in favour of $G_{Bab}$.

Renormalization must
now be addressed, and will be performed
in close analogy to the discussion in ~\cite{ditreu}.
The integral in eq.(~\ref{Gamma2scalB}) suffers from two
kinds of divergences: 

\begin{enumerate}

\item
An overall divergence of
the scalar proper-time integral $\int_0^{\infty}dT$
at the lower integration limit.

\item
Divergences of
the $\int_0^1\,du_a$ parameter integral 
at the points $0,1$ where the
endpoints of the photon propagator
become coincident, $u_a=u_b$.

\end{enumerate}
\noindent
The first one will be removed by 
one- and two-loop photon wave function
renormalization, the second one by one-loop 
scalar mass renormalization.
As is well known, vertex renormalization 
and scalar self energy renormalization
cancel against each other 
in this type of
calculation, and need not be considered. 

By power counting, an overall divergence can exist only for
the terms in the effective Lagrangian which are of order
at most quadratic in the external field $B$. 
Expanding the integrand of eq.(~\ref{Gamma2scalB}),
$K(z,u_a)\equiv {z\over\sinh(z)}A(z,u_a)$,
in the variable $z$, we find 

\begin{equation}
K(z,u_a)=
\biggl[
 {3\over 
{G_{Bab}}^2}-{12\over G_{Bab}}
\biggr]
+\biggl[
-{1\over 2}{1\over G_{Bab}^2}
+{1\over G_{Bab}}
+2
\biggr]
z^2
+O(z^4)
\label{scalintexp}
\end{equation}
\noindent
The complicated singularity appearing here at
the point $u_a=u_b$ indicates that this
form of the parameter integral is not yet
optimized for the purpose of
renormalization. In particular, it 
shows a spurious singularity 
in the coefficient of the
induced Maxwell term $\sim z^2$.
This comes not unexpected as the
cancellation of subdivergences implied
by the Ward identity has,
in a general gauge, 
no reason to be
manifest at the parameter
integral level.

We could improve on this either
by switching to Landau gauge, or
by performing a suitable partial
integration on the integrand.
The latter procedure is less systematic,
but easy enough to implement for the simple
case at hand:
Inspection of the two versions we
have of this parameter integral, 
the original one eq.(~\ref{Gamma2scal})
and the partially integrated
one eq.(~\ref{Gamma2scalpI}),
shows that we can optimize the integrand
by choosing a certain linear combination of
both versions, namely

\begin{equation}
{\cal L}^{(2)}_{\rm scal}[B]=
{3\over 4}\times
{\rm eq.}
(~\ref{Gamma2scal})
+{1\over 4}\times
{\rm eq.}
(~\ref{Gamma2scalpI})
\label{Gamma2scaloptim}
\end{equation}
\noindent
After integration over $\bar T$, this leads to
another version of eq.(~\ref{Gamma2scalB}),

\begin{equation}
{\cal L}^{(2)}_{\rm scal}
[B]=-
{(4\pi )}^{-4}
{e^2\over 2}
\int_{0}^{\infty}{dT\over T^3}e^{-m^2T}
{z\over \sinh(z)}
\int_0^1 du_a
\,A'(z,u_a)
\; ,
\label{Gamma2scalBprime}
\end{equation}
\noindent
with a different integrand

\begin{eqnarray}
A'&=&
\Biggl\lbrace
A'_0
{\ln ({G_{Bab}/ G_{Bab}^z})
\over{(G_{Bab}-G_{Bab}^z)}}
+
A'_1
{\ln ({G_{Bab}/ G_{Bab}^z})
\over{(G_{Bab}-G_{Bab}^z)}^2}
\nonumber\\
&&
+{A'_2\over
(G^z_{Bab})(G_{Bab}-G^z_{Bab})}
+{A'_3\over
(G_{Bab})(G_{Bab}-G^z_{Bab})}
\Biggr\rbrace\quad ,\nonumber\\
A'_0&=&
3\biggl[
2z^2G^z_{Bab}-{z\over\tanh(z)}-1
\biggr]
\quad , \nonumber\\
A'_1 &=&
A_1-{3\over 2}
\Bigl[
\dot G_{Bab}^2 -\dot G_{Bab}^{z2}
\Bigr]
\quad , \nonumber\\
A'_2&=&
A_2-{3\over 2}
\Bigl[
\dot G_{Bab}\dot G_{Bab}^z
+\dot G_{Bab}^{z2}
\Bigr]
\quad ,\nonumber\\
A'_3&=&
A_3+{3\over 2}
\Bigl[
\dot G_{Bab}^2
+\dot G_{Bab}\dot G_{Bab}^z
\Bigr]
\; . 
\nonumber\\
\label{A1A2A3prime}
\end{eqnarray}
\noindent
We have not yet taken into account here the term involving
$\delta_{ab}$, stemming from $\ddot{\cal G}_{Bab}$, which was
contained in the integrand of eq.(5.17). 
This term 
corresponds, in diagrammatic terms, to a tadpole insertion,
and could therefore be safely deleted. However, it will be
quite instructive to keep it and check explicitly that
it is taken care of by the renormalization procedure.
It leads to an integral $\int_0^{\infty} 
{d\bar T\over {\bar T}^2}$ which we regulate by introducing
an UV cutoff for the photon proper-time,

\be
\int_{{\bar T}_0}^{\infty}
{d\bar T\over {\bar T}^2}
={1\over {\bar T}_0}
\; .
\label{calctadpole}
\ee
\no
It gives then a further contribution $E({\bar T}_0)$ to
${\cal L}^{(2)}_{\rm scal}[B]$,

\begin{equation}
E({\bar T}_0)=-3
{(4\pi )}^{-4}
e^2
{1\over {\bar T}_0}
\int_{0}^{\infty}{dT\over T^2}e^{-m^2T}
{z\over \sinh(z)}
\label{tadpole}
\end{equation}
\no
Expanding the new integrand,
$K'(z,u_a)\equiv{z\over\sinh(z)}A'(z,u_a)$,
in $z$, we find the simple result

\begin{equation}
K'(z,u_a)=
-6{1\over G_{Bab}}
+3
z^2
+O(z^4)
\label{scalintexpprime}
\end{equation}
\noindent
In particular, the absence of a subdivergence for 
the Maxwell term is now manifest.

\noindent
We delete the irrelevant constant term, and add
and subtract the Maxwell term. 
Defining

\begin{equation}
K_{02}(z,u_a)=-6{1\over G_{Bab}}
+3
z^2,
\label{defK02}
\end{equation}
\no
the Lagrangian
then becomes

\begin{eqnarray}
{\cal L}^{(2)}_{\rm scal}
[B]&=&
E({\bar T}_0)
-{\alpha\over
2{(4\pi )}^{3}}
\int_{0}^{\infty}{dT\over T^3}e^{-m^2T}
\,3z^2\nonumber\\
&&
-{\alpha\over
2{(4\pi )}^{3}}
\int_{0}^{\infty}{dT\over T^3}e^{-m^2T}
\int_0^1 du_a
\,
\Bigl[
K'(z,u_a)-K_{02}(z,u_a)
\Bigr]
\;.\nonumber\\
\label{Gamma2scalBprime2}
\end{eqnarray}
\noindent
The second term, which we denote by $F$,
 is divergent when integrated over
the scalar proper-time
$T$. We regulate it
by introducing another
proper-time cutoff $T_0$
for the scalar proper-time
integral:

\be
F(T_0):=
-{\alpha\over
2{(4\pi )}^{3}}
\int_{2T_0}^{\infty}{dT\over T^3}e^{-m^2T}
\,3z^2
\label{defF}
\ee
\noindent
(we use $2T_0$ rather then $T_0$ for
easier comparison with ~\cite{ditreu}).
The third term is convergent at $T=0$, but
still has a divergence at $u_a=u_b$, as it
contains negative powers of
$G_{Bab}$. Expanding the integrand in
a Laurent series in $G_{Bab}$, one finds

\bear
K'(z,u_a)-K_{02}(z,u_a)&=&
{f(z)\over G_{Bab}}+O\Bigl(G_{Bab}^0\Bigr)
,\nonumber\\
f(z)&=&
3
\biggl[
2-{z\over\sinh(z)}-{z^2\cosh(z)\over\sinh(z)^2}
\biggr]
.\nonumber\\
\label{laurentGab}
\ear
\noindent
Again the singular part of this expansion
is added and subtracted, yielding

\begin{eqnarray}
{\cal L}^{(2)}_{\rm scal}
[B]&=&
E({\bar T}_0)
+
F(T_0)
-{\alpha\over
2{(4\pi )}^{3}}
\int_{2T_0}^{\infty}{dT\over T^3}e^{-m^2T}
\int_{T_0\over T}^{1-{T_0\over T}} du_a
\,{f(z)\over G_{Bab}}
\nonumber\\
&&
-{\alpha\over
2{(4\pi )}^{3}}
\int_{0}^{\infty}{dT\over T^3}e^{-m^2T}
\int_0^1 du_a
\,
\Bigl[
K'(z,u_a)-K_{02}(z,u_a)-{f(z)\over G_{Bab}}
\Bigr]
\; ,\nonumber\\
\label{Gamma2scalBprime3}
\end{eqnarray}
\noindent
The last integral is now completely finite.
The third term,
which we call $G(T_0)$,
is finite at $T=0$, as $f(z)=O(z^4)$
by construction. 
Here we have introduced
$T_0$ for the purpose of
regulating the divergence at $u_a=u_b$.
The $u_a$ -- integral for this term is then
readily computed and yields, in the limit
$T_0\rightarrow 0$, a contribution

\be
\int_{T_0\over T}^{1-{T_0\over T}}
du_a{1\over G_{Bab}}
=
-2{\rm ln}\Bigl({T_0\over T}\Bigr)
=-2{\rm ln}(\gamma m^2T_0)
+2{\rm ln}(\gamma m^2T).
\label{intua}
\ee
\noindent
We have rewritten this term for reasons
which will become apparent in a moment
($\ln (\gamma)$ denotes the Euler-Mascheroni
constant).
Next note that we can relate the function
$f(z)$ to the scalar one-loop Euler-Heisenberg
Lagrangian, eq.(~\ref{eulheiscal}).
If we write this Lagrangian for the pure
magnetic field case, and subtract the
two divergent
terms lowest order in $z$, 
we obtain

\be
\bar{\cal L}^{(1)}_{\rm scal}[B]
=
{1\over {(4\pi)}^2}\int_0^{\infty}
{dT\over T^3}{\rm e}^{-m^2T}
\biggl[
{z\over\sinh(z)}
+{z^2\over 6}-1
\biggr]
\label{Gamma1scalmagren}
\ee
\noindent
On the other hand, we can write

\bear
f(z)&=&
3
\biggl[
2-{z\over\sinh(z)}-{z^2\cosh(z)\over\sinh(z)^2}
\biggr]\nonumber\\
&=&
3T^3{d\over dT}
\biggl\lbrace
{1\over T^2}
\biggl[
{z\over\sinh(z)}+{z^2\over 6}-1
\biggr]
\biggr\rbrace
\nonumber\\
\label{rewritef}
\ear
\no
By a partial integration over $T$, we can therefore
reexpress

\bear
{1\over {(4\pi)}^2}
\int_0^{\infty}
{dT\over T^3}{\rm e}^{-m^2T}
f(z)
&=&
3{m^2\over {(4\pi)}^2}
\int_0^{\infty}
{dT\over T^2}{\rm e}^{-m^2T}
\biggl[
{z\over\sinh(z)}
+{z^2\over 6}-1
\biggr]\nonumber\\
&=&
-3m^2{\partial\over\partial m^2}
\bar{\cal L}^{(1)}_{\rm scal}[B]
\label{partint}
\ear
\no
To proceed, we need the value of the
one-loop mass displacement in scalar QED,
computed in the proper-time
regularization.
This we borrow from 
\cite{ginzburg}
\footnote
{Note that this differs by a sign from $\delta m^2$
as used in ~\cite{ss3}. Here this denotes the
mass displacement itself, there the corresponding counterterm.}:

\be
\delta m^2=
{3\alpha\over 4\pi}
m^2
\Bigl[
-{\rm ln}
\Bigl(\gamma m^2 T_0\Bigr)
+ {7\over 6}
+{1\over m^2{\bar T}_0}
\Bigr].
\label{deltamscal}
\ee
\no
Using this result, we may rewrite

\bear
G(T_0)&=&
\biggl[\delta m^2-{7\over 2}{\alpha\over 4\pi}m^2
-3{\alpha\over 4\pi {\bar T}_0}
\biggr]
{\partial\over\partial m^2}
\bar{\cal L}^{(1)}_{\rm scal}[B]
\nonumber\\
&&
-{\alpha\over {(4\pi)}^3}
\int_0^{\infty}
{dT\over T^3}{\rm e}^{-m^2T}
{\rm ln}(\gamma m^2 T)f(z)
.
\nonumber\\
\label{calcG}
\ear
\no
As expected the ${1\over {\bar T}_0}$ -- term
introduced by the one-loop mass renormalization
cancels the tadpole term $E({\bar T}_0)$,
up to its constant and Maxwell parts.
Moreover, the remaining divergence of
$G(T_0)$ for $T_0\rightarrow 0$
has been absorbed by $\delta m^2$.

Putting all pieces together, we can write
the complete two-loop
approximation to the
effective Lagrangian 
in the following way: 

\bear
{\cal L}_{\rm scal}^{(\le 2)}[B_0]
&=&
-\half B_0^2
-{1\over{(4\pi)}^2}
\int_{T_0}^{\infty}
{dT\over T^3}
{\rm e}^{-m^2_0T}
{z^2\over 6}
+
\bar{\cal L}^{(1)}_{\rm scal}[B_0]
+
\delta m_0^2
{\partial\over\partial m_0^2}
\bar{\cal L}^{(1)}_{\rm scal}
[B_0]\nonumber\\
&&-{7\over 2}{\alpha_0\over 4\pi}m_0^2
{\partial\over\partial m_0^2}
\bar{\cal L}^{(1)}_{\rm scal}
[B_0]
-{\alpha_0\over {(4\pi)}^3}
\int_0^{\infty}
{dT\over T^3}{\rm e}^{-m_0^2T}
{\rm ln}(\gamma m_0^2 T)f(z)
\nonumber\\
&&
-{\alpha_0\over
2{(4\pi )}^{3}}
\int_{0}^{\infty}{dT\over T^3}e^{-m_0^2T}
\int_0^1 du_a
\,
\Bigl[
K'(z,u_a)-K_{02}(z,u_a)-{f(z)\over G_{Bab}}
\Bigr]
\nonumber\\
&&
-{\alpha_0\over
2{(4\pi )}^{3}}
\int_{2T_0}^{\infty}{dT\over T^3}e^{-m_0^2T}
\,z^2
\Bigl(3-{T\over {\bar T}_0}\Bigr)
\nonumber\\
\label{Gammascalunrenorm}
\ear
We have rewritten this Lagrangian in
bare quantities, since up to now we have been
working in the bare regularized theory.
Only mass and photon
wave function renormalization are required
to make this effective Lagrangian finite:

\bear
m^2&=&m_0^2+\delta m_0^2,\nonumber\\
e&=&e_0Z_3^{\half},\nonumber\\
B&=&B_0Z_3^{-\half}.\nonumber\\
\label{renormalization}
\ear
\no
Here $\delta m_0^2$ has already been introduced
in eq.(~\ref{deltamscal}), while $Z_3$ is
chosen such as to absorb the diverging
one- and two-loop Maxwell terms in eq.
(~\ref{Gammascalunrenorm}).
Note that this leaves $z=e_0B_0T$ unaffected.
The final answer becomes
\footnote{\noindent
Note added in proof: The constant $7\over 2$
multiplying the third term is incorrect, and should be
replaced by $9\over 2$. This has now been established
both by a detailed comparison with \cite{ritusscal}, and
another recalculation using
dimensional regularization
\cite{frss}.} 

\bear
{\cal L}_{\rm scal}^{(\le 2)}[B]
&=&
-\half B^2
+
{1\over {(4\pi)}^2}\int_0^{\infty}
{dT\over T^3}{\rm e}^{-m^2T}
\biggl[
{z\over\sinh(z)}
+{z^2\over 6}-1
\biggr]
\nonumber\\
&&
+{7\over 2}{\alpha\over {(4\pi)}^3}m^2
\int_0^{\infty}
{dT\over T^2}{\rm e}^{-m^2T}
\biggl[
{z\over\sinh(z)}
+{z^2\over 6}-1
\biggr]
\nonumber\\
&&
-{\alpha\over
2{(4\pi )}^{3}}
\int_{0}^{\infty}{dT\over T^3}e^{-m^2T}
\int_0^1 du_a
\,
\Bigl[
K'(z,u_a)-K_{02}(z,u_a)-{f(z)\over G_{Bab}}
\Bigr]
\nonumber\\
&&
-{\alpha\over {(4\pi)}^3}
\int_0^{\infty}
{dT\over T^3}{\rm e}^{-m^2T}
{\rm ln}(\gamma m^2 T)f(z)
\nonumber\\
\;.
\label{Gammascalrenorm}
\ear
This parameter integral representation is
of a similar but simpler structure than the
one given by Ritus ~\cite{ritusscal}. 

The corresponding result
for the case of a pure electric field is obtained
from it by the simple substitution
(in Minkowski space)

\be
B\rightarrow -iE.
\label{BtoE}
\ee
\no
This makes an important
and well-known difference.
In the 
electric field case the 
$T$ -- integration
acquires new divergences
due to the appearance of poles.
This leads to an imaginary part
of the effective action, 
and to a probability for
electron-positron pair creation.
At the one-loop level, the first
such pole becomes significant
at the critical
field strength

\be
E_{cr}={m^2\over e}
\label{Ecrit}
\ee\no
(see eq.(~\ref{eulheiscal})).
The two-loop contribution
to the effective Lagrangian
affects also the imaginary part
and the pair creation probability.
A detailed investigation of those
corrections has been undertaken in
\cite{lebedev,ginzburg}.

The calculation for the
case of a generic constant field
would be only moderately more difficult, 
if one uses the Lorentz frame
where the magnetic and electric fields
are parallel, as in ~\cite{ritusscal}.

\vskip40pt

{\bf 6. The Two-Loop Euler-Heisenberg Lagrangian
for Spinor QED}
\renewcommand{\theequation}{6.\arabic{equation}}
\setcounter{equation}{0}
\vskip15pt
The corresponding
calculation for the spinor loop case proceeds
in complete analogy when formulated in the
superfield formalism. This allows us to
immediately write down the analogue of 
eqs.(~\ref{Gamma2scal}),(
\ref{wickscal}):
\begin{eqnarray}
{\cal L}^{(2)}_{\rm spin}
[F]&=&
(-2)
{(4\pi )}^{-D}
\Bigl(-{e^2\over 2}\Bigr)
\int_0^{\infty}{dT\over T}e^{-m^2T}T^{-{D\over 2}} 
\int_0^{\infty}d\bar T 
\int_0^T d\tau_a d\tau_b
\int d\theta_a d\theta_b
\nonumber\\
&\phantom{=}&\times
{\rm det}^{-{1\over 2}}
\biggl[{\tan(eFT)\over {eFT}}
\biggr]
{\rm det}^{-{1\over 2}}
\biggl[
\bar T 
-{1\over 2}
\hat{\cal C}_{ab}
\biggr]
\langle
-D_ay_a\cdot D_by_b\rangle
\quad ,
\nonumber\\
\label{Gamma2spin}
\end{eqnarray}
\no
with a superfield Wick contraction

\begin{equation}
\langle
-D_ay_a\cdot D_b y_b\rangle
=
{\rm tr}
\biggl[
D_aD_b\hat{\cal G}_{ab}+{1\over 2}
{D_a( \hat{\cal G}_{aa}- \hat{\cal G}_{ab})
D_b( \hat{\cal G}_{ab}- \hat{\cal G}_{bb})
\over
{\bar T -{1\over 2}\hat{\cal C}_{ab}}}
\biggr]\; .
\nonumber\\
\label{superwick}
\end{equation}
The notations should be obvious.

Performing the Grassmann integrations 
$\int d\theta_a \int d\theta_b$, and
removing $\ddot {\cal G}_{Bab}$ by partial integration as
before, we obtain the equivalent of 
eq.(~\ref{Gamma2scalpI}),

\vfill\eject

\begin{eqnarray}
{\cal L}^{(2)}_{\rm spin}[F]&=&
{(4\pi )}^{-D}
e^2
\int_0^{\infty}{dT\over T}e^{-m^2T}T^{-{D\over 2}} 
\int_0^{\infty}d\bar T 
\int_0^T d\tau_a
\int_0^T d\tau_b
\nonumber\\
&\phantom{=}&\times
{\rm det}^{-{1\over 2}}
\biggl\lbrack{\tan(eFT)\over {eFT}}
\Biggr\rbrack
{\rm det}^{-{1\over 2}}
\biggl\lbrack
\bar T -{1\over 2}
{\cal C}_{ab}
\biggr\rbrack
\nonumber\\
&&
\times
{1\over 2}
\Biggl\lbrace
{\rm tr}\dot{\cal G}_{Bab}
{\rm tr}
\biggl\lbrack
{\dot{\cal G}_{Bab}
\over
{\bar T -{1\over 2}{\cal C}_{ab}}}
\biggr\rbrack
-{\rm tr}{\cal G}_{Fab}
{\rm tr}\biggl[
{{\cal G}_{Fab}\over
{\bar T -{1\over 2}{\cal C}_{ab}}}
\biggr]
\nonumber\\
&\phantom{=}&
+{\rm tr}
\biggl[
{(\dot {\cal G}_{Baa}-\dot {\cal G}_{Bab})
(\dot {\cal G}_{Bab}-\dot {\cal G}_{Bbb}+2{\cal G}_{Faa})
+{\cal G}_{Fab}{\cal G}_{Fab}
-{\cal G}_{Faa}{\cal G}_{Fbb}
\over
{\bar T -{1\over 2}{\cal C}_{ab}}}
\biggr]
\Biggr\rbrace\; .
\nonumber\\
\label{Gamma2spinpI}
\end{eqnarray}
\noindent
In writing this formula, we have used the symmetry between
$\tau_a$ and $\tau_b$ to reduce the number of terms.
The same expressions could have been obtained starting from
eq.(~\ref{Gamma2scalpI}) and using the generalized one-loop
substitution rule.

Specializing to the pure magnetic case
and $D=4$, it is then again a
matter of simple algebra to calculate the traces and $\bar T$
-- integrals. 
After rescaling and setting $\tau_b=0$, 
the result can be written as

\begin{equation}
{\cal L}^{(2)}_{\rm spin}
[B]=
{\alpha\over{(4\pi )}^{3}}
\int_{T_0}^{\infty}{dT\over T^3}e^{-m^2T}
{z\over \tanh(z)}
\int_0^1 du_a
B(z,u_a),
\label{Gamma2spinB}
\end{equation}
\noindent
with

\begin{eqnarray}
B(z,u_a)&=&
\Biggl\lbrace
B_1
{\ln ({G_{Bab}/G_{Bab}^z})
\over{(G_{Bab}-G_{Bab}^z)}^2}
+{B_2\over
G^z_{Bab}(G_{Bab}-G^z_{Bab})}
+{B_3\over
G_{Bab}(G_{Bab}-G^z_{Bab})}
\Biggr\rbrace,\nonumber\\
B_1&=&A_1-4z\tanh(z)G^z_{Bab}
\quad , \nonumber\\
B_2&=&A_2+8z\tanh(z)G^z_{Bab}-3
\quad ,\nonumber\\
B_3&=&A_3-4z\tanh(z)G^z_{Bab}+3
\quad .
\nonumber\\
\label{B1B2B3}
\end{eqnarray}
\noindent
Comparison with an earlier field theory calculation 
performed by
Dittrich and one of the authors ~\cite{ditreu} shows that 
the integrand of eq.(~\ref{Gamma2spinB})
allows for a direct identification with its counterpart
there, as given in eqs. (7.21),(7.22). This
requires nothing more than
a rotation to Minkowskian proper-time,
$T\rightarrow is$, a transformation of variables
from $u_a$ to $v:=\dot G_{Bab}$, and the
use of trigonometric identities.
In particular,
our quantities $G_{Bab},G_{Bab}^z$ then identify with
the quantities $a,b$ there.

The renormalization of this Lagrangian has,
for the spinor-loop case, been
carried through in detail in that work.
We will therefore not repeat this analysis
here, and just give the final result for the
renormalized two-loop contribution
to the Euler-Heisenberg Lagrangian
\footnote{Note added in proof: The constant
$-10$ multiplying the third term is
incorrect, and should be replaced
by $-18$ \cite{frss}
(compare the footnote before 
eq.(\ref{Gammascalrenorm})).}
: 

\bear
{\cal L}_{\rm spin}^{(\le 2)}[B]
&=&
-\half B^2
-
{2\over {(4\pi)}^2}\int_0^{\infty}
{dT\over T^3}{\rm e}^{-m^2T}
\biggl[
{z\over\tanh(z)}
-{z^2\over 3}-1
\biggr]
\nonumber\\
&&
-10m^2
{\alpha\over {(4\pi)}^3}\int_0^{\infty}
{dT\over T^2}{\rm e}^{-m^2T}
\biggl[
{z\over\tanh(z)}
-{z^2\over 3}-1
\biggr]
\nonumber\\
&&
+{2\alpha\over {(4\pi)}^3}
\int_0^{\infty}
{dT\over T^3}{\rm e}^{-m^2T}
{\rm ln}(\gamma m^2 T)g(z)
\nonumber\\
&&
+{\alpha\over
{(4\pi )}^{3}}
\int_{0}^{\infty}{dT\over T^3}e^{-m^2T}
\int_0^1 du_a
\,
\Bigl[
L(z,u_a)-L_{02}(z,u_a)-{g(z)\over G_{Bab}}
\Bigr]\nonumber\\
\label{Gammaspinrenorm}
\ear\no
with

\bear
L(z,u_a)&=&{z\over\tanh(z)}B(z,u_a),\nonumber\\
L_{02}(z,u_a)&=&-{12\over G_{Bab}}+2z^2,\nonumber\\
g(z)&=&-6\biggl[
{z^2\over{\sinh(z)}^2}+z\coth(z)-2
\biggr].\nonumber\\
\label{defLLg}
\ear\no
For a study of the strong field limit of this
Lagrangian see again ~\cite{ditreu}.

The first exact calculation of this
two-loop Lagrangian for the
fermion loop case is again due
to Ritus ~\cite{ritusspin},
who also used proper-time methods
to arrive at a certain two-parameter integral.

The integral representation
given above is equivalent to the
one given by Ritus, but simpler.  
In \cite{ditreu} it was obtained 
by convoluting a free photon propagator with the
polarization tensor of a fermion in a
constant magnetic field. The essential part of
this calculation consists of deriving a compact integral representation
for the polarization tensor. To this end, complicated expressions
involving Dirac traces and momentum integrals have to be evaluated.
In the string-inspired case, no analogous manipulations are needed,
and the computational effort for doing the parameter integrals
which it introduces instead is much smaller.

Moreover, when applied to spinor QED, the method of the present paper
yields the corresponding result for scalar QED with almost no
further effort. This would not be the case for the standard field
theory techniques.

Concerning the physical relevance of this type of calculation,
let us mention the experiment PVLAS in preparation at Legnaro,
Italy, which
is an optical experiment designed to yield the first experimental
measurement of the Euler-Heisenberg Lagrangian
~\cite{bakalovetal,bakalov}. It is conceivable that the
technology used there may even allow for the measurement of
the two-loop correction in the near future ~\cite{bakalov,zavattini}. 

\vfill\eject

{\bf 7. The 2-Loop QED $\beta$ -- Functions Revisited}
\renewcommand{\theequation}{7.\arabic{equation}}
\setcounter{equation}{0}
\vskip10pt

Finally, let us remark that the method we have employed
in this paper for the calculation of the full two-loop
Euler-Heisenberg Lagrangians also improves on the
calculation of the two-loop QED $\beta$ -- functions
as it had been presented in ~\cite{ss3}. For the
extraction of the $\beta$ -- function coefficients
one needs only to calculate the induced Maxwell terms.
Up to the contributions from one-loop mass 
renormalization,
the correct two-loop scalar 
~\cite{bialni} and spinor \cite{joslut}
QED coefficients
can thus be read
off from the expansions
(see 
eq.(~\ref{scalintexpprime}),
eq.(~\ref{defLLg}))

\bear
K'(z,u_a)
&=&
-{6\over G_{Bab}}+3z^2+O(z^4),\nonumber\\
L(z,u_a)
&=&
-{12\over G_{Bab}}+2z^2+O(z^4)
.\nonumber\\
\label{spinintexpprime}
\ear\no
Comparing with ~\cite{ss3} we see that the
use of the generalized Green's functions ${\cal G}_B,
{\cal G}_F$ has saved us two integrations: 
The same formulas
eq.(~\ref{master}) which there had been employed
for executing the integrations over the 
points of interaction
$\tau_1,\tau_2$ with the external field,
have now entered already at the level of the
construction of those Green's functions.
Of course, for the $\beta$ -- function calculation
all terms of order higher than $O(F^2)$ are irrelevant,
so that one could then as well use the
truncations of those Green's functions given in
eq.(~\ref{GB(F)expand}). Moreover, one would choose
an external field with the property $F^2\sim \bf I$.

Note that in the fermion loop
case a subdivergence-free integrand
was obtained proceeding directly from the
partially integrated version eq.(~\ref{Gamma2spinpI}).
This fact, which had already been noticed in ~\cite{ss3},
is not accidental, and can be understood by an analysis
of the quadratic divergences. In the scalar QED case, there are
three possible sources of quadratic divergences for the
induced Maxwell term:

\begin{enumerate}
\item
The contact term containing $\delta_{ab}$.

\item
The leading order term $\sim {1\over G_{Bab}^2}z^2$
in the ${1\over G_{Bab}}$ -- expansion
of the main term (see e.g. eq.(\ref{scalintexp})).

\item
The explicit $1/{\bar T}_0$ appearing in the one--loop 
mass displacement eq.(~\ref{deltamscal}).

\end{enumerate}
\no
The last one should cancel the other two in the
renormalization procedure, if those are regulated by the
same UV cutoff ${\bar T}_0$ for the photon proper-time,
and this was verified
in various versions of this calculation.
In the spinor QED case the fermion propagator has no quadratic 
divergence (this is, of course, manifest in the first order
formalism, while in the second order formalism there are various
diagrams contributing to the one--loop fermion self energy, and the
absence of a quadratic divergence is due to a cancellation among
them). The third term is thus missing, and the other two have to
cancel among themselves. In particular, the completely partially
integrated version of the integrand has no $\delta_{ab}$ -- term
any more, and consequently the second term must also be absent.
However the ${1\over G_{Bab}}$ -- expansion
of the main contribution to the Maxwell 
term is, if one does this calculation in $D=4$, 
always of the form
shown in eq.(\ref{scalintexp}),

\be
\biggl[
{A\over G_{Bab}^2}+{B\over G_{Bab}}+C
\biggr]{\rm tr}(F^2)
\label{maxwellexp}
\ee
\no
with coefficients $A,B,C$.
In the partially integrated version first consideration of the
quadratic subdivergence allows one to conclude that $A=0$, and then
consideration of the logarithmic subdivergence that $B=0$. 

Note that this argument does not apply
to the scalar QED case, nor does it to
spinor QED in dimensional
renormalization, due to the suppression of quadratic divergences by
that scheme. In both cases one would have only one constraint
equation for the two coefficients $A$ and $B$ appearing in the
partially integrated integrand, and indeed
they turn out to be nonzero in both cases.
In the present formalism, the fermion
QED two-loop $\beta$ -- function calculation 
thus becomes simpler when performed
not in dimensional regularization, but
in some four-dimensional scheme such as proper-time
or Pauli-Villars regularization.
 
The reader may rightfully ask why we have gone
to such lengths in analyzing this 2-loop calculation,
which is easy to do by modern standards 
even in field theory. 
We find this cancellation mechanism
interesting in
view of some facts known about the 
three-loop 
fermion QED $\beta$ -- function
\cite{rosner,brandt,bdk}.
Apart from the well-known cancellation
of transcendentals occuring between diagrams
in the calculation of the
quenched (one fermion loop) contribution
to this $\beta$ -- function \cite{rosner,bdk},
which takes place in any scheme and gauge, even more
spectacular cancellations were found in 
~\cite{brandt} where this
calculation was performed in 
four dimensions,
Pauli-Villars regularization, and
Feynman gauge. 
In that calculation all contributions 
from
nonplanar diagrams happened to cancel out
exactly.
A recalculation of this coefficient in the
present formalism is currently being undertaken
~\cite{fss2}.

\vskip15pt

{\bf 8. Discussion}
\renewcommand{\theequation}{7.\arabic{equation}}
\setcounter{equation}{0}
\vskip10pt
In this paper, we have extended previous work of two of the
authors 
on the multiloop generalization of the string-inspired
technique to the case of quantum electrodynamics in a constant
external field.  
The resulting formalism has been tested
on a recalculation of the two-loop corrections to the
Euler-Heisenberg Lagrangian for quantum electrodynamics.
Several advantages of this calculus over standard field theory
methods have been pointed out.
In particular, it treats the scalar and spinor loop
cases on the same footing, so that the scalar loop results are
always obtained as a byproduct of the 
corresponding spinor loop calculations.
More technically,
our parameter integrals are written in a form
convenient for partial integrations.
In particular, the integrands are 
functions well-defined on the circle, so that boundary
terms do not appear. The usefulness of
this property has been demonstrated in the
renormalization of the scalar QED
two-loop Lagrangian.

An application to a recalculation
of the one-loop QED photon splitting amplitude 
has been given separately ~\cite{adlsch}.

We have derived a path integral
representation of the gluon loop, 
and used it for a recalculation
of what is the closest analogon to the one-loop
Euler-Heisenberg Lagragian in Yang-Mills theory.
More significantly,
the analysis of chapter 3 should be viewed as a first step towards
an extension of the worldline technique to multiloop
calculations in nonablian gauge theory. 
Our derivation of
this path integral is entirely non-heuristic, 
and thus guaranteed to reproduce the correct one-loop off-shell
amplitudes for Yang-Mills theory. 
From our experience with quantum electrodynamics,
this property alone makes us optimistic
about the existence of a multiloop
generalization of the worldline method
for the Yang-Mills case. 
We hope to have more to say about this
in the future.

\vskip15pt
{\bf Acknowledgements:}
We would like to thank S. L. Adler,
Z. Bern, D. Broadhurst, A. Denner, W. Dittrich and E. Zavattini
for various discussions and informations.
C. Schubert also thanks the 
Deutsche Forschungsgemeinschaft
for financial support, and the 
Institute for 
Advanced Study, Princeton,
for hospitality during 
the final stage of this project.

\vfill\eject

{\bf Appendix A: Path Integral Representation of the 
Electron Propagator}

\renewcommand{\theequation}{A.\arabic{equation}}
\setcounter{equation}{0}
\vskip10pt
\noindent

Different from the case of the closed loop fermionic 
worldline Lagrangian eq.(~\ref{spinorpi}),
for the Dirac propagator besides the einbein field gauged
to $T$ one has to introduce a gravitino field $\chi$ in the world line
action. This can be gauged to a constant, but not to zero. For
massive fermions a further field $\psi_5(\tau)$ coupling to mass has to be
introduced ~\cite{bdzdh,bdh,polbook} (its supersymmetric partner $x_5$ is
not needed for the gauge coupling, but
essential for the worldline implementation of Yukawa couplings 
~\cite{mnss1}).

Integration  over $\chi$ in the path integral
leads to a factor  $(-\frac{1}{2}\psi_\mu\dot x_\mu+m\psi_5)$
corresponding to the numerator
of the Dirac propagator. This can be demonstrated nicely
for the free propagator ~\cite{bc,hht,holten}
in the  coherent state formalism. 
This ``holomorphic representation''
represents operators in terms of their (fermionic) Wick symbols.

In the case with background interaction considered here we prefer
a different approach in which fermionic operators are represented
by their {\it Weyl} symbols.
This formalism is manifestly covariant and, contrary to the
holomorphic representation, it treats propagators in external
fields and one-loop effective actions on the same footing. From
a canonical point of view we are dealing with the following
algebra of hermitian operators $\hat\psi_\mu$ and
$\hat\psi_5$:
\bear\label{A.1}
&&\hat\psi_\mu\hat\psi_\nu+\hat\psi_\nu\hat\psi_\mu=\delta_{\mu\nu},
\nonumber\\
&&\hat\psi_\mu\hat\psi_5+\hat\psi_5\hat\psi_\mu=0,
\ \ \  \psi^2_5=\frac{1}{2}\ear
In terms of euclidean Dirac matrices, it can be represented
by
\be\label{A.2}
\hat\psi_\mu=\frac{i}{\sqrt 2}\gamma_5\gamma_\mu,\quad
\hat\psi_5=\frac{1}{\sqrt 2}\gamma_5\ee
The Weyl symbol map ``symb'' establishes a linear one-to-one
map between operators and functions of the anticommuting $c$-numbers
$\xi_\mu$ and $\xi_5$. In particular, $\symb(\hat\psi_\mu)
=\xi_\mu$ and $\symb(\hat\psi_5)
=\xi_5$.  The inverse symbol map associates the Weyl-ordered (totally
antisymmetrized) operator product to strings of $\xi$'s:
\be\label{A.3}
\symb^{-1}(\xi_\mu\xi_\nu...\xi_\rho)=\{\hat\psi_\mu\hat\psi_\nu...
\hat\psi_\rho\}_{\rm Weyl}\ee
For example,
\be\label{A.4}
\symb^{-1}(\xi_\mu\xi_\nu)=\frac{1}{2}
(\hat\psi_\mu\hat\psi_\nu-\hat\psi_\nu\hat\psi_\mu)=
\hat\psi_\mu\hat\psi_\nu-\frac{1}{2}\delta_{\mu\nu}\ee
and $\symb(\hat\psi_\mu\hat\psi_\nu)=\xi_\mu\xi_\nu+\frac{1}{2}
\delta_{\mu\nu}$. (See ref. [103,104] for further details.)

Let us consider the Dirac propagator in the background of an arbitrary
abelian gauge field and let us write down a path integral
representation for its kernel (bosonic variables)/symbol (fermionic
variables)
\be\label{A.5}
G^{\rm Dirac}(x_2,x_1;\xi)=\symb\Big[\langle x_2|(\slD+m)^{-1}
|x_1\rangle\Big](\xi)\ee
with $\slD\equiv\gamma_\mu D_\mu=2i\hat\psi_\mu\hat\psi_5D_\mu$.
After having integrated out the auxiliary fields $\chi$
and $x_5$ it reads [13,27,104], up to
an overall constant:
\vfill\eject
\bear\label{A.6}
&&G^{\rm Dirac}(x_2,x_1;\xi)\propto\int^T_0dT e^{-m^2T}\\
&&\ \ \ \times \int^{x(T)=x_2}_{x(0)=x_1}{\cal D}x\quad
\int_{\psi(0)+\psi(T)=2\xi}{\cal D}\psi_\mu\quad
\int_{\psi_5(0)+\psi_5(T)=2\xi_5}{\cal D}\psi_5\nonumber\\
&&\times\frac{1}{T}\int^T_0d\tau\{-\frac{1}{2}
\psi_\mu(\tau)\dot x_\mu(\tau)
+m\psi_5(\tau)\}\ \exp[-S_B-S_F-S_5]\nonumber\ear
The action consists of the following pieces:
\bear\label{A.7}
&&S_B=\int^T_0 d\tau [\frac{1}{4}\dot x^2_\mu+ieA_\mu(x)\dot x_\mu]
\nonumber\\
&&S_F=\int^T_0 d\tau [\frac{1}{2}\psi_\mu\dot\psi_\mu-ie F_{\mu\nu}(x)
\psi_\mu\psi_\nu]+\frac{1}{2}\psi_\mu(T)\psi_\mu(0)\nonumber\\
&&S_5=\int_0^T d\tau \frac{1}{2}\psi_5\dot\psi_5+\frac{1}{2}
\psi_5(T)\psi_5(0)\ear
Note the surface terms in $S_F$ and $S_5$. They are needed in
order to correctly reproduce the equations of motion [104].
The factor $\frac{1}{T}\int^T_0d\tau\{...\}$ in eq. (\ref{A.6})
stems from the integration over the world line gravitino field
$\chi$. It is important to realize that the terms inside
the curly brackets are actually independent of $\tau$,
and that we may replace
\be\label{A.8}
\frac{1}{T}\int^T_0d\tau\{-\frac{1}{2}
\psi_\mu(\tau)\dot x_\mu(\tau)+m\psi_5(\tau)\}
\longrightarrow -\frac{1}{2}\psi_\mu(T)\dot x_\mu(T)+m\psi_5(T)\ee
The $\tau$-independence of the expectation value of $\psi_\mu
\dot x_\mu$ is a consequence of the supersymmetry of the
action. In fact, $S_B+S_F$ is invariant under
\bear\label{A.9}
&&\delta x_\mu=-2\eta\psi_\mu\nonumber\\
&&\delta\psi_\mu=\eta\dot x_\mu\ear
with a {\it constant} Grassmann parameter $\eta$. If, instead,
a time-dependent parameter is used in (\ref{A.9}),
the action changes according to
\be\label{A.10}
\delta(S_B+S_F)=\int^T_0d\tau \ \eta(\tau) \ \frac{d}{d\tau}
(\psi_\mu\dot x_\mu)\ee
Obviously $\psi_\mu\dot x_\mu$ is the conserved
Noether charge related to the supersymmetry (\ref{A.9}).
If we apply a localized supersymmetry transformation
to the path integral $\int{\cal D}x{\cal D}\psi
\exp(-S_B-S_F)$ and observe that the measure is
invariant we obtain the Ward identity
\be\label{A.11}
\frac{d}{d\tau}\langle \psi_\mu(\tau)\dot x_\mu(\tau)
\rangle =0\ee
Eq. (\ref{A.11}) together with a similar argument for
$\psi_5$ justifies the replacement (\ref{A.8}).

Using (\ref{A.8}) in (\ref{A.6}), the insertion $-\frac{1}{2}\psi
_\mu\dot x_\mu+m\psi_5$ is evaluated at the final point of
the trajectory, $\tau =T$. Hence it may be pulled in front
of the path integral, then acting as a (differential/matrix) operator
on the wave function which was time-evolved by the path integral.
If we are dealing with a phase-space path integral of the
type
\be\label{A.12}
\int^{x(T)=x_2}_{x(0)=x_1}{\cal D}x\int {\cal D} p\ \exp
\{\int^T_0d\tau(ip\dot x-H)\}\ee
we know that
\bear      &&\int{\cal D}x{\cal D}p \ x(T)\ \exp\{...\}=x_2\
\int{\cal D}x{\cal D}p\  \exp\{...\}     \\
&&    \int{\cal D}x{\cal D}p \  p(T)\ \exp\{...\}=-i\frac{\partial}{
\partial x_2}\int{\cal D}x{\cal D}p\ \exp\{...\}    \ear
By rewriting eq. (\ref{A.6}) in hamiltonian form,
it is easy to see that in the case at hand $\dot x_\mu(T)$
corresponds to the operator $-2iD_\mu(x_2)$ acting from
the left. With an analogous reasoning for the fermions
this leads us to the following representation for the
Dirac propagator:
\bear\label{A.13}
&&G^{\rm Dirac}(x_2,x_1)\propto[\hat\psi_\mu iD_\mu(x_2)+m\hat\psi_5]\\
&&\ \ \ \ \times\int^\infty_0dTe^{-m^2T} \
    \symb^{-1}\Big[K^{\rm Dirac}(x_2,T|x_1,0;\xi_
\mu)\  I_5(\xi_5,T)\Big]\nonumber\ear
Here we defined
\be\label{A.14}
K^{\rm Dirac}(x_2,T|x_1,0;\xi_\mu)\equiv
\int^{x(T)=x_2}_{x(0)=x_1}{\cal D}x\int_{\psi(0)+
\psi(T)=2\xi}{\cal D}\psi_\mu \ e^{-S_B-S_F}\ee
and
\be\label{A.15}
I_5(\xi_5,T)\equiv\int_{\psi_5(0)+\psi_5(T)=2\xi_5}{\cal D}
\psi_5  \  e^{-S_5}\ee
Eq. (\ref{A.13}) was obtained from (\ref{A.6})
by applying the inverse symbol map. As for the fermionic
degrees of freedom, $G^{\rm Dirac}(x_1,x_2)$ is an operator
now, i.e., a matrix acting on spinor indices.

                                              Up to this
point, no assumption about the gauge field $A_\mu(x)$
has been made. From now on we consider fields with
$F_{\mu\nu}=const.$ In this case the path integral (\ref{A.14})
factorizes:
\be\label{A.16}
K^{\rm Dirac}(x_2,T|x_1,0;\xi_\mu)=
K_B(x_2,T|x_1,0) \ I_F(\xi_\mu,T)\ee
The bosonic piece
\be\label{A.17}
K_B(x_2,T|x_1,0)=\int^{x(T)=x_2}_{x(0)=x_1}{\cal D}
x\ e^{-S_B}\ee
is the same as in the spin-0 or spin-1 case. Its evaluation
is described in detail in section 3. The result is given
by (3.68) with $g$ replaced by $e$, and with the
factor $\exp[2igTF]_{\mu\nu}$ omitted. What remains to be
done is to calculate the fermionic contribution
\bear\label{A.18}
I_F(\xi_\mu,T)&=&\int_{\psi(0)+\psi(T)=2\xi}{\cal D}
\psi_\mu\ \exp[-\frac{1}{2}\psi_\mu(T)\psi_\mu(0)]\nonumber\\
&&\ \   \times\exp\left\{-\frac{1}{2}\int^T_0d\tau\ 
\psi_\mu[\partial_{\tau}
\delta_{\mu\nu}-2ie F_{\mu\nu}]\psi_\nu\right\}\ear
Since the $\psi$-integral is Gaussian, the saddle point method
will yield its exact value. We decompose the integration variable
according to
\be\label{A.19}
\psi_\mu(\tau )=\psi_\mu^{\rm class}(\tau )+\varphi_\mu(\tau )\ee
where $\psi_\mu^{\rm class}$ is a solution of the classical
equation of motion,
\be\label{A.20}
[\partial_{\tau}\delta_{\mu\nu}-2ie F_{\mu\nu}]
\psi_\nu^{\rm class}(\tau )=0,\ee
subject to the boundary condition $\psi_\mu^{\rm class}(0)+
\psi_\mu^{\rm class}(T)=2\xi_\mu$. The fluctuation field
$\varphi_\mu$ satisfies antiperiodic boundary conditions.
Using (\ref{A.19}) and (\ref{A.20}) in (\ref{A.18}) we obtain
\bear\label{A.21}
I_F(\xi_\mu,T)&=&\exp[-\frac{1}{2}\psi_\mu^{\rm class}(T)
\psi_\mu^{\rm class}(0)]\nonumber\\
&&\times\int_A{\cal D}\varphi\ \exp\left\{-\frac{1}{2}\int^T_0
d\tau\  \varphi_\mu[\partial_{\tau}
\delta_{\mu\nu}-2ie F_{\mu\nu}]\varphi_\nu
\right\}\nonumber\\
&=&\exp[-\frac{1}{2}\psi_\mu^{\rm class}(T)\psi_\mu^{\rm class}
(0)]\ {\rm Det}_A[\partial_{\tau}
\delta_{\mu\nu}-2ieF_{\mu\nu}]^{1/2}\ear
In analogy with section 3, the determinant in (\ref{A.21}) is
given by $2^D\det_L [\cos(eFT)]$. The solution to
(\ref{A.20}) which satisfies the correct boundary conditions reads
\be\label{A.22}
\psi_\mu^{\rm class}(\tau )=2\left(\frac{\exp
(2ieF\tau )}{1+\exp(2ieFT)}\right)_{\mu\nu}\xi_\nu\ee
Inserting this into (\ref{A.21}) leads us to the final
result for $I_F$:
\bear\label{A.23}
I_F(\xi_\mu,T)&=&2^{D/2}\ {\rm det}_L[\cos(eFT)]^{1/2}\nonumber\\
&&\ \times\exp[i\xi_\mu\tan(eFT)_{\mu\nu}\xi_\nu]\ear
Using the same method we can show that $I_5$ equals an unimportant
constant which we shall drop. Thus, because
\be\label{A.24}
\symb^{-1}(K^{\rm Dirac})=K_B\ \symb^{-1}(I_F),\ee
our last task is to find out which is the operator corresponding
to the symbol (\ref{A.23}). We shall see that
\bear\label{A.25}
&&\symb[\exp(ieTF_{\mu\nu}\hat\psi_\mu\hat\psi_\nu)](\xi)\nonumber\\
&&\ \  \ ={\rm det}_L[\cos(eFT)]^{1/2}\  \exp[i\xi_\mu\tan(eFT)_{\mu\nu}
\xi_\nu]\ear
In order to prove (\ref{A.25}), we transform $F_{\mu\nu}$ to
block-diagonal form {We assume $D$ even.}
\be\label{A.26}
F_{\mu\nu}={\rm diag}  \left[{
0\ ~~B_1\choose-B_1\ 0},\    {
0\ ~~B_2\choose -B_2\ 0},\    \cdots    \right]\ee
Since the $\hat\psi_\mu$'s pertaining to different blocks
are mutually anticommuting, we may prove (\ref{A.25})
for each block separately. Focusing on the first one,
it is convenient to define
\be\label{A.27}
\hat\Sigma_{12}\equiv i(\hat\psi_1\hat\psi_2-\hat\psi_2\hat\psi_1)
\ee
As this operator is Weyl-ordered, we have
\be\label{A.28}
\symb   \left[ \hat\Sigma_{12} \right]   =2i\xi_1\xi_2\ee
Because $(\hat\Sigma_{12})^2=1$, it follows that
\bear\label{A.29}
\exp[ieTF_{\mu\nu}\hat\psi_\mu\hat
\psi_\nu]&=&\exp[eB_1T\hat\Sigma_{12}]\nonumber\\
&=&\cosh(eB_1T)-\hat\Sigma_{12}\sinh(eB_1T)\ear
and therefore
\bear\label{A.30}
&&\symb[\exp(ieTF_{\mu\nu}\hat\psi_\mu\hat\psi_\nu)]
(\xi)\nonumber\\
&&\ \ \ \ \ \  = \cosh(eB_1T)[1-2i\xi_1\xi_2\tanh(eB_1T)]\nonumber\\
&&\ \ \ \ \ \  = {\rm det}_L[\cos(eFT)]^{1/2}\exp[i\xi_\mu\tan(eFT)
_{\mu\nu}\xi_\nu]\ear
In the last line of (\ref{A.30}) we used that the eigenvalues
of the first block are $\pm iB_1$ and that
$\xi_1^2=0=\xi_2^2$. Repeating the same argument
for the other blocks establishes eq. (\ref{A.25}).

Upon inserting (\ref{A.24}) with (\ref{A.23}) and
(\ref{A.25}) into (\ref{A.13}), we obtain the well-known
result for the euclidean Dirac propagator in a constant
background field [104,105]:
\bear\label{A.31}
G^{\rm Dirac}(x_1,x_2)&=&[-\gamma_\mu D_\mu(x_2)+m]
\int^\infty_0dT(4\pi T)^{-D/2}e^{-m^2T}\nonumber\\
&&\ \cdot \exp\left[-\frac{1}{4}(x_2-x_1)eF\cot(eTF)(x_2-x_1)
\right]\nonumber\\
&&\ \cdot  \exp\left[-\frac{1}{2}\tr_L\ln\frac{\sin(eTF)}{(eTF)}\right]
\nonumber\\
&&\ \cdot \exp\left[+\frac{i}{2}eTF_{\mu\nu}\gamma_\mu\gamma_\nu\right]
\ear
In (\ref{A.31}) we used the representation (\ref{A.2})
for the $\hat\psi$'s, and we dropped an overall factor of
$\gamma_5$ which is produced by the path integral, but is
not included in the standard definition of $G^{\rm Dirac}$.
The expression for the bosonic contribution $K_B$ was taken
from section 3. It applies to the Fock-Schwinger gauge centered
at $x_1$. In the general case, $K_B$ contains an extra phase
factor $\exp(-ie\int^{x_2}_{x_1}dx_\mu A_\mu)$.
We also note that the scalar propagator
$(-D^2)^{-1}$
is obtained from (\ref{A.31}) by simply deleting the operator
$[-\slD + m]$
and the last exponential involving the $\gamma$-matrices.

It is remarkable that the above calculation of the
propagator is almost identical to the calculation
of the one-loop effective action,
the only difference being the boundary
condition of the fermionic path integral. For the propagator
we need $\psi_\mu(T)+\psi_\mu(0)=2\xi_\mu$, whereby
the variables $\xi_\mu$ give rise to its
$\gamma$-matrix structure. The effective action, on the
other hand, is a scalar quantity, and it is obtained from the
same path integral with $\xi_\mu=0$. Giving a non-zero value
to $\xi_\mu$ amounts to creating a fermion line by ``opening''
a loop.

\vfill\eject
{\bf Appendix B: Derivation of Worldline
Green's Functions}

\renewcommand{\theequation}{B.\arabic{equation}}
\setcounter{equation}{0}
\vskip10pt
\noindent
The worldline Green's functions 
appearing in this paper
are kernels of certain integral operators, acting in 
the real Hilbert space of periodic or
antiperiodic functions defined on an interval
of length $T$. 
We denote by $\bar H_P$ the full space of periodic 
functions, by $H_P$ the same space with the constant
mode exempted, and by $H_A$ the space of antiperiodic
functions. The ordinary derivative acting on those
functions is correpondingly denoted by
$\partial_P$, $\bar\partial_P$ or $\partial_A$.
With those definitions, we can write our Green's functions
as

\begin{eqnarray}
{\cal G}_B(\tau_1,\tau_2)&=&2
\langle \tau_1\mid
{\Bigl(
{\partial_P}^2
-2iF\partial_P
\Bigr)}^{-1}
\mid
\tau_2\rangle,\nonumber\\
{\cal G}_F(\tau_1,\tau_2)&=&2
\langle \tau_1\mid
{\Bigl(
{\partial_A}
-2iF\Bigr)}^{-1}\mid
\tau_2\rangle,\nonumber\\
{\cal G}_P^C(\tau_1,\tau_2)&=&
\langle \tau_1\mid
{\Bigl(
{\bar\partial_P}-C
\Bigr)}^{-1}
\mid \tau_2\rangle,\nonumber\\
{\cal G}_A^C(\tau_1,\tau_2)&=&
\langle \tau_1\mid
{\Bigl(
{\partial_A}-C\Bigr)}^{-1}
\mid \tau_2\rangle.\nonumber\\
\label{collectGreen's}
\end{eqnarray}
\noindent
(in this appendix we 
absorb the coupling constant
$e$ into the external field $F$).
Note that ${\cal G}^C_A$ is,
up to a conventional factor of 2,
formally identical with
${\cal G}_F$ under the replacement $C\rightarrow 2iF$.

${\cal G}_B$ and ${\cal G}_F$
are easy to construct
using the following representation of the 
integral kernels for inverse
derivatives on the unit circle ~\cite{ss3}

\begin{eqnarray}
\langle u
\mid {{\partial}_P}^{-n}\mid
u'\rangle
&=&
-{1\over n!}B_n(\mid u-u'\mid)
{\rm sign}^n(u-u')\nonumber\\
\langle u
\mid {{\partial}_A}^{-n}\mid
u'\rangle&=&
{1\over{2(n-1)!}}
E_{n-1}(\mid u-u'\mid)
{\rm sign}^n(u-u' )\quad .\nonumber\\
\label{master}
\end{eqnarray}
\noindent
Here $B_n$($E_n$) denotes the
$n$-th Bernoulli (Euler) polynomial.
Those formulas are valid for $\mid u-u'\mid \le 1$.
For instance, the computation of ${\cal G}_B$ 
proceeds as follows:


\begin{eqnarray}
{\cal G}_B(u_1,u_2)&=&
2\langle u_1\mid {\Bigl(
{\partial_P}^2
-2iF\partial_P
\Bigr)}^{-1}\mid u_2\rangle
\nonumber\\
&=&2
\sum_{n=0}^{\infty}{(2iF)}^n
\langle u_1\vert
{\partial_P}^{-(n+2)}
\vert u_2\rangle\nonumber\\
&=&
-2\sum_{n=2}^{\infty}
{{(2iF)}^{n-2}{{\rm sign}^n(u_1-u_2)}
\over n!}
B_n(\mid u_1 -u_2\mid )\nonumber\\
&=&-{1\over iF}
{{\rm sign}(u_1-u_2)
{\rm e}^{2iF(u_1-u_2)}
\over
{{\rm e}^{2iF{\rm sign}(u_1-u_2)}
-1}}
+{{\rm sign}(u_1-u_2)\over iF}
B_1(\mid u_1-u_2\mid ) -{1\over 2F^2}\nonumber\\
&=& {1\over 2F^2}\biggl({F\over{{\rm sin}F}}
{\rm e}^{-iF\dot G_{B12}}+iF\dot G_{B12} -1\biggr)
\quad .
\nonumber\\
\label{calGB}
\end{eqnarray}
\noindent
In the next-to-last step we used the
generating identity for the
Bernoulli polynomials,

\begin{equation}
{t{\rm e}^{xt}\over{{\rm e}^t-1}}=
\sum_{n=0}^{\infty}  
B_n(x){t^n\over n!}.
\label{defbernoullipol}
\end{equation}
\noindent
This is ${\cal G}_B$ as given in
eq.(\ref{calGBGF}) up to a simple
rescaling.
The computation of ${\cal G}_F$
proceeds in a completely analogous way.

This method does not work for the determination of
${\cal G^C_P}$, as negative powers of
$\bar\partial_P$ are not even well-defined
in the presence of the zero mode.

In the following, we will calculate ${\cal G^C_{P,A}}$
in a different, more
``physical'' way,
which corresponds to the usual construction of 
the Feynman propagator in
field theory.

In order to determine ${\G}^C_A(\tau)$, say, we employ the
following set of basis functions over the circle with
circumference $T$:
\be\label{B.1}
f_n(\tau)=T^{-1/2}\exp[i\frac{2\pi}{T}(n+\frac{1}{2})\tau],\ \
n\in Z\ee
They satisfy
\bear\label{B.2}
&&\int_0^Td\tau f_n^*(\tau)f_m(\tau)=\delta_{nm},\nonumber\\
&&\sum_{n=-\infty}^\infty f_n(\tau_2)f_n^*(\tau_1)=
\sum_{m=-\infty}^\infty\delta(\tau_2-\tau_1-mT)\ear
and $f_n(\tau+T)=-f_n(\tau)$. In this basis, the Green's function
(\ref{3.28}) becomes
\be\label{B.3}
{\cal G}^C_A(\tau_1-\tau_2)=\frac{1}{T}\sum^\infty_{n=-\infty}
\frac{\exp[i\frac{2\pi}{T}(n+\frac{1}{2})(\tau_1-\tau_2)]}
{i(2\pi/T)(n+\frac{1}{2})-C}\ee
By introducing an auxiliary integration in the form $(\tau\equiv\tau
_1-\tau_2)$
\be\label{B.4}
{\cal G}_A^C(\tau)=\int^\infty_{-\infty}d\omega\ \frac{1}{T}
\sum^\infty_{n=-\infty}\delta
\biggl(\omega-\frac{2\pi}{T}(n
+\frac{1}{2})\biggr)\frac{e^{i\omega\tau}}{i\omega-C}\ee
and using Poisson's resummation formula, the Green's
function assumes
the suggestive form \cite{kleinert}
\be\label{B.5}
{\G}_A^C(\tau)=\sum^\infty_{n=-\infty}(-1)^n{\G}_\infty^C(\tau+nT)\ee
with
\be\label{B.6}
{\G}^C_\infty(\tau)\equiv\int^\infty_{-\infty}
\frac{d\omega}{2\pi}
\frac{e^{i\omega\tau}}{i\omega-C}\ee
We verify that
\be\label{B.7}
[\partial_\tau-C]{\G}_\infty^C(\tau)=\delta(\tau)\ee
\be\label{B.8}
[\partial_\tau-C]{\G}^C_A(\tau)=\sum^\infty_{n=-\infty}
(-1)^n\delta(\tau+nT)\ee
which shows that ${\G}^C_\infty$ is a Green's function on the
infinitely extended real line, while ${\G}^C_A$ is defined
on the circle. The integral (\ref{B.6}) yields for $C>0$
\be\label{B.9}
{\G}^C_\infty(\tau)=-\Theta(-\tau)e^{C\tau}\ee
Hence, from (\ref{B.5})
\be\label{B.10}
{\G}_A^C(\tau)=-e^{C\tau}\sum_{n=-\infty}^\infty(-1)^n
\Theta(-\tau-nT)e^{nCT}\ee
For $\tau\in (0,T)$ only the terms $n=-\infty,...,-1$
contribute to the sum in (\ref{B.10}), while for $\tau
\in(-T,0)$ a nonzero contribution is obtained for $n=-\infty,
...,0$. Summing up the geometric series in either case
and combining the results we obtain the expression given
in eq. (\ref{3.29}). It is valid for
$-T<\tau<+T$. Using a basis of periodic functions the same
arguments lead to ${\G}^C_P$ as stated in (\ref{3.29}).
Note that in the limit of a large period $T$
\be\label{B.11}
\lim_{T\to\infty}{\cal G}^C_{A,P}(\tau)={\G}^C_\infty(\tau),\ee
as it should be. For $C\to 0$, both ${\G}^C_\infty$ and
${\G}^C_A$ have a well-defined limit:
\bear\label{B.12}
&&{\G}^0_\infty(\tau)=-\Theta(-\tau)\nonumber\\
&&{\G}^0_A(\tau)=\frac{1}{2}\ {\rm sign}\ (\tau)\ear
The periodic Green's function ${\G}^C_P$ blows up in this limit
because $\bar\partial_P^{-1}$ does not exist in presence of the
constant mode. It is important to keep in mind that ${\G}^C_P$
is defined in such a way that it includes the zero mode of
$\partial_\tau$.

In the perturbative evaluation of the spin-1 path integral one has
to deal with traces over chains of propagators of the form
\be\label{B.13}
\sigma_{A,P}^n(C)\equiv
{{\Tr}}_{A,P}[(\partial_\tau-C)^{-n}]\ee
Because
\be\label{B.14}
\sigma_{A,P}^n(C)=\frac{1}{(n-1)!}\left(\frac{d}{dC}\right)^{n-1}
\sigma_{A,P}^1(C)\ee
it is sufficient to know $\sigma_{A,P}^1(C)$. The subtle point which
we would like to mention here is that strictly speaking
the sum defining $\sigma_A^1$, say,
\be\label{B.15}
\sigma_A^1(C)=\sum^\infty_{n=-\infty}\frac{1}{i(2\pi/T)(n+\frac{1}
{2})-C}\ee
does not converge as it stands, and it is meaningless without
a prescription of how to regularize it. The usual strategy
is to combine terms for positive and negative values of $n$,
and to replace (\ref{B.15}) by the convergent series
\bear\label{B.16}
\sigma_A^1(C)&=&-2C\sum^\infty_{n=0}\left[\left(\frac{2\pi}
{T}\right)^2(n+\frac{1}{2})^2+C^2\right]^{-1}\nonumber\\
&=&-\frac{T}{2}\tanh\left(\frac{CT}{2}\right).\ear
It is important to realize that this definition implies
a well-defined prescription for the treatment of
the $\Theta$ functions in ${\G}^C_{A,P}$ at $\tau=0$. In
fact,
\be\label{B.17}
\sigma_A^1(C)=\int^T_0d\tau\ {\G}^C_A(\tau-\tau)=T\ {\G}^C_A(0),
\ee
and by combining eqs. (\ref{B.16}) and (\ref{B.17}) we deduce
that we must set
\be\label{B.18}
\lim_{\tau\searrow0}\ \Theta(\tau)=\lim_{\tau\searrow0}
\ \Theta(-\tau)=\frac{1}{2}.\ee
With (\ref{B.16}) we obtain
\be\label{B.19}
\sigma_A^n(C)=-\frac{1}{(n-1)!}\left(\frac{T}{2}\right)^n
\left(\frac{d}{dx}\right)^{n-1}\tanh(x)\Bigr|_{x=CT/2}.\ee
The analogous relation in the periodic case is
\be\label{B.20}
\sigma^n_P(C)=-\frac{1}{(n-1)!}\left(\frac{T}{2}\right)^n
\left(\frac{d}{dx}\right)^{n-1}\coth(x)\Bigr|_{x=CT/2}\ee
if the zero mode of $\partial_\tau$ is included
in the trace (\ref{B.13}), and
\be\label{B.21}
{\sigma'}^n_P(C)=-\frac{1}{(n-1)!}\left(\frac{T}{2}\right)^n
\left(\frac{d}{dx}\right)^{n-1}\{\coth(x)-x^{-1}
\}\Bigr|_{x=CT/2}\ee
if the zero mode is omitted. For $C$ sufficiently small
one finds the power series expansions
\bear\label{B.22}
&&\sigma_A^n(C)=-\frac{1}{(n-1)!}\sum^\infty_{k=n/2}
\frac{(2^{2k}-1)B_{2k}}{2k(2k-n)!}T^{2k}C^{2k-n}\nonumber\\
&&{\sigma'}_P^n(C)=-\frac{1}{(n-1)!}\sum^\infty_{k=n/2}
\frac{B_{2k}}{2k(2k-n)!}T^{2k}C^{2k-n}\ear
$\sigma_A^n$ and ${\sigma'}_P^n$ have well defined limits
for $C\to0$:
\bear\label{B.23}
&&\sigma_A^n(0)=-\frac{(2^n-1)B_n}{n!}T^n
=\half{E_{n-1}\over (n-1)!}T^n
\quad (n {\rm\; even})
\nonumber\\
&&{\sigma'}_P^n(0)=-\frac{B_n}{n!}T^n
\quad (n {\rm\; even})
\ear
\noindent
Those limits vanish for $n$ odd.
This brings us, of course, back to eqs.(\ref{master}).

\vfill\eject
{\bf Appendix C: Worldline Determinants}
\vskip10pt

\renewcommand{\theequation}{C.\arabic{equation}}
\setcounter{equation}{0}
\vskip10pt
\noindent
In this appendix we collect a few results about the
determinants which arise in the computation of the
spin-1 wordline path integral. To start with, we consider
the operator $\partial_\tau+\omega$ where $\omega$ is a
real constant, and $\partial_\tau$ acts on periodic
and antiperiodic functions of period $T$, respectively.
Its spectrum reads $i(2\pi/T)n$ in the former and
$i(2\pi/T)(n+1/2)$ in the latter case, $n\in Z$. Ratios of
determinants of the form
\be\label{C.1}
\frac{{{\Det}}_A[\partial_\tau+\omega]}{{{\Det}}_A[\partial_\tau]}
=\prod^\infty_{n=-\infty}\frac{i(\frac{2\pi}{T})(n+\frac{1}{2})
+\omega}{i(\frac{2\pi}{T})(n+\frac{1}{2})}\ee
are defined by the prescription that terms with positive and
negative values of $n$ should be combined so as to obtain
the manifestly convergent product
\be\label{C.2}
\frac{{{\Det}}_A[\partial_\tau+\omega]}{{{\Det}}_A[\partial_\tau]}
=\prod^\infty_{n=0}\left[1+\left(\frac{\omega T}{2\pi}\right)^2
\frac{1}{(n+1/2)^2}\right]=\cosh\left(\frac{\omega T}{2}\right)\ee
In the periodic case we omit the zero mode from the definition
of the determinants and find likewise
\be\label{C.3}
\frac{{\Det}_P'[\partial_\tau+\omega]}{{\Det}_P'[\partial_\tau]}
=\frac{\sinh(\omega T/2)}{(\omega T/2)}\ee
(compare eqs.(~\ref{scaldetcomp}), (~\ref{grassfermcomp})).
Next we look at the matrix differential operator

$$(\partial_\tau-C)\delta_{\mu\nu}+\Omega_{\mu\nu},\  \mu,
\nu=1,...,D.$$

\noindent
Here $\Omega$ is a constant matrix. We assume
that it can be diagonalized and has eigenvalues $\omega$.
For antiperiodic boundary conditions we obtain from (\ref{C.2})
\be\label{C.4}
\frac{{\Det}_A[(\partial_\tau-C)\delta_{\mu\nu}+\Omega
_{\mu\nu}]}{{\Det}_A[(\partial_\tau-C)\delta_{\mu\nu}]}
=\prod_\omega\frac{\cosh[\frac{T}{2}(C-\omega)]}{
\cosh[CT/2]}.\ee
The product extends over the spectrum of $\Omega$. The corresponding
formula for periodic boundary conditions, with the zero mode
removed, reads
\be\label{C.5}
\frac{{\Det}_P'[(\partial_\tau-C)\delta_{\mu\nu}+\Omega
_{\mu\nu}]}{{\Det}'_P[(\partial_\tau-C)\delta_{\mu\nu}]}
=\prod_{\omega'}\frac{C}{C-\omega'}\ \prod_\omega
\frac{\sinh[\frac{T}{2}(C-\omega)]}{
\sinh[CT/2]}.\ee
For $C\not=0$ we can reinstate the zero mode of $\partial_\tau$.
In this case (\ref{C.5}) is replaced by
\be\label{C.6}
\frac{{\Det}_P[(\partial_\tau-C)\delta_{\mu\nu}+\Omega
_{\mu\nu}]}{{\Det}_P[(\partial_\tau-C)\delta_{\mu\nu}]}
=\prod_\omega\frac{\sinh[\frac{T}{2}(C-\omega)]}{
\sinh[CT/2]}.\ee

In this paper we use the above determinants for an exact
evaluation of the worldline path integral in the background of a
constant field $F_{\mu\nu}$. For more complicated field
configurations only a perturbative calculation of
the path integral is possible in general. It is based upon
the Green's functions ${\G}_{A,P}^C$ which were discussed
in appendix B. It can be checked that the determinants
given above are consistent with the perturbative expansion.
In perturbation theory, the l.h.s. of eq. (\ref{C.4}), for
instance, is interpreted as a power series in $\omega$:
\bear\label{C.7}
&&\prod_\omega{\Det}_A[1+\omega{\G}^C_A]=\prod_\omega
\exp{\Tr}\ln[1+\omega{\G}^C_A]\nonumber\\
&&=\prod_\omega\exp\left\{-\sum_{n=1}^\infty\frac{(-1)^n}
{n}\omega^n\sigma^n_A(C)\right\}\ear
In the last line of (\ref{C.7}) we have used (\ref{B.13}).
By virtue of eq. (\ref{B.19}) one can sum up the
perturbation series in closed form:
\bear\label{C.8}
&&-\sum^\infty_{n=1}\frac{(-1)^n}{n}\omega^n\sigma^n_A(C)
\nonumber\\
&&\quad=\sum_{n=1}^\infty(-1)^n\frac{(\omega T)^n}{2^nn!}\left
(\frac{d}{dx}\right)^{n-1}\tanh(x)\Bigr|_{x=CT/2}\nonumber\\
&&\quad=\left[\exp\left(-\frac{\omega T}{2}\frac{d}{dx}\right)-1\right]
\ln\cosh(x)\Bigr|_{x=CT/2}\nonumber\\
&&\quad=\ln\frac{\cosh[x-\omega T/2]}{\cosh(x)}\Bigr|_{x=CT/2}\ear
With (\ref{C.8}) inserted into (\ref{C.7}) we reproduce
precisely the r.h.s. of eq. (\ref{C.4}).

\vfill\eject

\end{document}